\documentclass[12pt,a4paper]{article}
\usepackage[a4paper, total={7.4in, 9.2in}]{geometry}
\usepackage[utf8]{inputenc}
\usepackage{amsmath}
\usepackage{amsfonts}
\usepackage{amssymb}
\usepackage{graphicx}
\usepackage[round, sort]{natbib}
\usepackage[colorlinks=true, linkcolor=blue, citecolor=blue, urlcolor=blue, pdfborder={0 0 0}]{hyperref}
\usepackage{leftindex}

\title{Streamlining business functions in official statistical production with Machine Learning\footnote{The views expressed in this contribution are those of the authors and do not necessarily reflect the views of Statistics Spain (INE).}}

\author{Sandra Barragán \and Adrián Pérez-Bote \and Carlos Sáez \and David Salgado \and Luis Sanguiao-Sande\\[1em]
Statistics Spain (INE)
}
\date{\today}

\begin{document}

\maketitle

\abstract{We provide a description of pilot and production experiences to streamline some business functions in the official statistical production process using statistical learning models. Our approach is quality-oriented searching for an improvement on accuracy, cost-efficiency, timeliness, granularity, response burden reduction, and frequency. Pilot experiences have been conducted with data from real surveys in Statistics Spain (INE).}

\section{The production of official statistics, the new data ecosystem, Artificial Intelligence, and quality}
\label{sec:ProdOffStat}
In the last two decades the production of official statistics has increasingly been under stronger pressure to deliver \textquotedblleft wider, deeper, quicker, better, cheaper\textquotedblright\ statistical products \citep{Hol07a, Bra07a, Nor07a}. This increasing pressure has grown along with the evolution of the economic, scientific, and technological progress of our society \citep[see e.g.][Chapter~1]{Rad20a}.\\

Interestingly enough, the first electronic digital computer for civil usage produced in the United States of America was firstly devoted to the production of official statistics at the U.S. Census Bureau \citep{USBurCen24a}. Nowadays, the computational power of statistical offices is far from that of, say, top IT companies. Before the Internet age, statistical offices used to concentrate comparatively vast amount of data. In the last decade, we are living an increasing data deluge \citep{Gle11a} and one of the most outstanding challenges for statistical offices is to access and use this data to reduce response burden, to gain efficiency, and to increase quality in the production of official statistics \citep{DGINS13a, DGINS18a}. More recently, the overwhelming deployment of applications based on Artificial Intelligence and Machine Learning (AI/ML) techniques clearly reveals the breed of experts excelling in the use of data and statistical models outside the community of official statistics. The incorporation of Data Science, Machine Learning, and Artificial Intelligence into the routinely production of official statistics is a current top priority in the modernization of statistical offices \citep{UNE21a}.\\

In our view, the two main ingredients in this pressing environment are the new data ecosystem and the exponentially progressive success of Artificial Intelligence and Machine Learning, i.e.\ the use of new data sources and new statistical methods to increase quality in many of its dimensions. On the one hand, the new data ecosystem arises as a natural consequence of the digitization driving the big data phenomenon and of the central role of data in the digital economy bringing the need of data governance, data management, and data stewardship in the construction of data spaces and similar data ecosystems. The integration of all kind of digital transactional and administrative data together with survey data into the production of official statistics stands as a natural demand on official statistics \citep[see e.g.][]{Han18a}.\\

On the other hand, the public explosion of Artificial Intelligence applications naturally poses the question on its use by statistical officers. This is currently an intense activity in official statistics understanding not only off-the-peg applications but especially the adaptation of statistical and deep learning models, in particular in the realm of finite population estimation problems \citep[see contribution by][in this same volume]{PutSalDaa24a}.\\

This combination of factors (data + AI/ML), in our view, constitutes a devilishly complex challenge for official statistics. Traditionally, the discipline of official statistics has been working with so-called microdata, i.e.\ basically data matrices of multiple variables per statistical unit (households, establishments, enterprises, etc.). The digitization of the last decades is increasingly bringing into play so-called nano\-data, i.e. transactional data with a much finer degree of entity breakdown. This is the data incrementally feeding more and more AI/ML systems already impinging on what and how statistical offices produce (or should now produce), and impacting on the role of official statistics in society. To take an example, if policymakers have nowadays access to this nanodata and data scientists can process and analyse it \citep[see e.g.][]{GueMar24a}, what is the role of official statistics produced by statistical offices?\\

Regarding the first factor, the adaptation to the new data ecosystem must be undertaken beyond doubt, with the subsequent transformation and modernization of multiple data management aspects \citep[see e.g.][]{DAMA17a}. However, data and statistics must not be confused \citep{Rei23a} and the main mission of official statistics, i.e.\ to provide a quantitative description of the connection between data and reality, including an uncertainty assessment (statistical inference), must prevail, all under a scrupulous fulfillment of legal regulations of statistical products (release calendar, accuracy, cost restrictions, territorial and sectorial breakdowns, data privacy and statistical confidentiality, etc.). In this renewed scenario, a revamped quality management and quality assurance must play a central role embracing old and new aspects providing knowledge and a standard reference for an increasingly datafied society \citep[see e.g.][for far-reaching consequences regarding new data sources]{FT25a}.\\

Regarding the second factor, by and large, we distinguish two broad approaches in the use of AI/ML techniques in the production of official statistics. Firstly, as in any other industry, any task or activity related to information-processing should be subjected to an assessment for the potential adoption of an AI/ML tool increasing the cost efficiency, improving timeliness, and enhancing overall quality. This is the case e.g. of statistical classification coding (with automatic coding systems), dissemination (e.g. with a chatbot), computer code production (with copilots and generative AI), etc. Secondly, however, as a specificity of the business of official statistical production, statistical inference must be paid due attention since it constitutes the critical core of the whole business. Regarding the first approach, the identification, adaptation and adoption of AI/ML tools is just a matter of time and resource investment to come up with the best options. Regarding the second approach, which is indeed deeply connected to our conception of statistical quality (beyond data quality), it will require greater efforts: is design-based inference still the preferred choice for inference? How estimators should be improved with these new models? Do we need to change the inference paradigm? Are all underlying statistical assumptions in AI/ML valid for the inference problem in official statistics?\\

Here we present ongoing initiatives at Statistics Spain (INE) to use statistical learning models to improve the production of official statistics. Our approach is quality-oriented, where we focus on critical aspects of statistical products such as timeliness, granularity, accuracy, efficiency, response burden, and frequency and seek improvements in different business functions using the versatile statistical learning models. For this chapter we shall adopt the definition of business function provided by the Generic Statistical Information Model (GSIM) v2.0 \citep[][]{GSIMv20}: "Activities undertaken by a statistical organisation to achieve its objectives".\\

The chapter is organized as follows. In Section~\ref{sec:StrTradBusFunc} we describe how to possibly improve core business functions such as estimation, editing, and classification coding in a more traditional fashion. In Section~\ref{sec:NewBusFunc} we propose alternative uses of these models to execute novel business functions to improve different quality dimensions. In Section~\ref{sec:Concl} we close with some conclusions.

\section{Streamlining traditional business functions}
\label{sec:StrTradBusFunc}

We propose improved business functions related to inference, editing, and classification coding.

\subsection{Design-based predictive inference}
\label{sub:AlgAssEst}
The official statistics approach to inference has been traditionally the design-based inference from probability samples \citep[see e.g.][]{Han87a, Smi94a,  Kal02a, Rao05a}. This approach relies on a \emph{known} sampling design and thus it is valid by construction. The usual alternative approach to inference, is model-based inference, where the uncertainty of estimation is evaluated with respect to an \emph{assumed} statistical model. Model-based inference often leads to more accurate results but might be invalid because of model misspecification.\\ 

In order to recover (at least part of) this loss of efficiency, a model-assisted approach has been proposed \citep{SSW92, WuSit01a, BO17}, where an auxiliar model is explicitly formulated but inference remains design-based. Model-assisted estimators are often not design-unbiased, but design-consistent asymptotically for a hypothetical sequence of populations of increasing sizes. However, design-unbiased model-assisted estimators have been also proposed \citep[see][]{HR54, Mic59, SZ20}.\\ 

In this section we present a brief summary of the work by \citet{ZSL24}, where estimation is given by an arbitrary prediction algorithm, while the uncertainty measure (mean squared error) is design-based. This separation between estimator and its properties is what makes this approach different from any model-assisted estimation.\\

Denote by $U = \{ 1, ...., N\}$ a given finite population of size $N$. Let $y_k, k\in U$, be the associated values of interest. Denote by $\mathbf{x}_k, k\in U$, the collection of feature vectors, where $\mathbf{x}_k$ is the vector associated with each unit $k\in U$. Given any sample of units from $U$, denoted by $s\subset U$, let $\mu(\mathbf{x}, s)$ denote the prediction given the feature vector $\mathbf{x}$ for a certain prediction algorithm trained on the sample $s$. Note that the algorithm can be model-based or just any heuristic Machine Learning algorithm. When the target parameter is the population total $Y = \sum_{k\in U} y_k$, the prediction estimator of $Y$ is given as
\begin{equation} \label{predEst}
\hat{Y} = \sum_{k\in s} y_k + \sum_{k\in U\setminus s} \mu(\mathbf{x}_k, s).
\end{equation}

Note that many design-based estimators in survey sampling can as well be given as prediction estimators. Of course, as $\mu$ is arbitrary, this prediction estimator is in general biased, so variance is not a good measure of its accuracy and we need to estimate both bias and mean squared error. Unfortunately, it is not possible to measure the bias for the model trained on the full sample, since we lack out-of-sample units to compare with model predictions. This has already been noticed too for usual Machine Learning cross-validation \citep{BHT23}.\\

Following the ideas by \citet{SZ20}, denote by $s_1 \cup s_2 = s$ and $s_1 \cap s_2 = \emptyset$ a \emph{training-test sample split}, where $s_1$ is selected by a \emph{subsampling design}, denoted by $q(s_1 \mid s)$, and the sample $s$ is selected according to a sampling design $p(s)$. Denote by $\mu(\mathbf{x}, s_1)$ the predictor obtained from the subsample $s_1$, in the same way as $\mu(\mathbf{x}, s)$ from $s$. Its error $\mu(\mathbf{x}_k, s_1) - y_k$ can be observed for any $k\in s_2$.\\

We shall refer to the sampling design that yields $(s_1, s)$ as the \emph{$pq$-design}, denote by 
\begin{equation} \label{pg-design}
f_{pq}(s_1, s) = q(s_1 \mid s) p(s) = f(s \mid s_1) f(s_1) 
\end{equation} 
where the last product indicates that, conditional on the training set $s_1$, one can view the test set $s_2$ as a probability sample from $U\setminus s_1$, according to which $s$ can vary under the $pq$-design. In particular, for any $k\in U$, let 
\begin{equation} \label{pi2}
\pi_{2k} = \mathbb{P}(k\in s_2 \mid s_1)  = \sum_{s\ni k, k\notin s_1} f(s\mid s_1)
\end{equation} 
be its conditional $s_2$-inclusion probability given $s_1$ under the $pq$-design.\\

Now, for a given subsample $s_1$ under the $pq$-design, the \emph{subsample-trained} prediction estimator is
\[
\hat{Y}_1^* = \sum_{k\in s} y_k + \sum_{k\in U\setminus s} \mu(\mathbf{x}_k, s_1)
\]
Note that while $\mu$ is trained on the subsample, its predictions are used out of the full sample $s$. Now, applying Rao-Blackwellisation to $\hat{Y}_1^*$ we obtain the subsampling Rao-Blackwellised (SRB) prediction estimator:
\begin{equation} \label{RBest}
\hat{Y}^{RB} = \sum_{k\in s} y_k + \sum_{k\in U\setminus s} \bar{\mu}(\mathbf{x}_k, s), 
\end{equation}
where 
\begin{equation} \label{SRB}
\bar{\mu}(\mathbf{x}_k, s) = \mathbb{E}_q\left[ \mu(\mathbf{x}_k, s_1) \mid s\right]
\end{equation}

As we point out below, this conditional expectation value in practice is computed using Monte Carlo approximations.\\

Note that this SRB prediction estimator is still a prediction estimator but for a slightly different algorithm $\bar{\mu}$. This new predictor is trained on the full sample. However, we cannot be sure whether it is more accurate or not than the estimator $\mu$ trained with the full sample. The advantage of using $\bar{\mu}$ over $\mu$ trained with the full sample is that it allows to get design-based estimators for both its bias and its MSE.\\

Let us denote $e_{1k} = \mu(\mathbf{x}_k, s_1) - y_k$ for any $k\notin s_1$, then 
\begin{equation} \label{biasRB}
\hat{B}^{RB} = \mathbb{E}_q(\sum_{k\in s_2} ( \pi_{2k}^{-1} - 1) e_{1k})
\end{equation}
is a (design-unbiased) estimator of the bias.\\

The construction of an MSE estimator is not so straightforward, but as shown by \citet[][Theorem~1]{ZSL24}, a design-unbiased estimator for the MSE of the SRB prediction estimator is:
\begin{equation} \label{mseRB}
\mbox{mse}^{RB} = \mathbb{E}_q\{ \hat{B}^2 - \hat{V}_s(\hat{B} \mid s_1) + \hat{V}_s\{ B(s_2) \mid s_1\} \mid s\} - V_q(\hat{Y}_1^* \mid s) 
\end{equation}
where $\hat{B} = \sum_{k\in s_2} ( \pi_{2k}^{-1} -1) \{ \mu(\mathbf{x}_k, s_1) - y_k \}$, and $\hat{V}_s(\hat{B} \mid s_1)$ is unbiased for
\[
V_s(\hat{B} \mid s_1) = \sum_{k\notin s_1} \sum_{l\notin s_1} (\pi_{2kl} - \pi_{2k} \pi_{2l}) 
\left(\frac{1}{\pi_{2k}} -1\right) \left(\frac{1}{\pi_{2l}} -1\right) e_{1k} e_{1l}
\]
where $\pi_{2ij} = \mathbb{P}(i, j\in s_2 \mid s_1)$, and $\hat{V}_s\{ B(s_2) \mid s_1\}$ is unbiased for
\[
V_s\{ B(s_2) \mid s_1\} = \sum_{k\notin s_1} \sum_{l\notin s_1} (\pi_{2kl} - \pi_{2k} \pi_{2l}) e_{1k} e_{1l} ~.
\]

Regarding the computation of the MSE estimator, note that:
\begin{enumerate}
    \item The number of subsamples for exact Rao-Blackwellisation is $\binom{n}{n_1}$ where $n_1$ is the subsample size. Since each subsample requires to fit the model again, exact Rao-Blackwellisation is usually computationally intractable.
    \item Thus, in practice, we have to replace exact RB with Monte-Carlo RB by approximating conditional expectations by sample means. More concretely, we choose some positive integer $T \leq \binom{n}{n_1}$ and approximate $\mathbb{E}_q\left[ \mu(\mathbf{x}_k, s_1)\right]$ by $\frac{1}{T}\sum_{t=1}^T \mu(\mathbf{x}_k, s_1^{(t)})$ based on sample splits $(s_1^{(t)}, s_2^{(t)})$. We similarly approximate the conditional expectations appearing in $\mbox{mse}^{RB}$ by a sample mean.
    \item This increases the variance of both $\hat{Y}^{RB}$ and $\mbox{mse}^{RB}$.
    \item While the Monte-Carlo variance of $\hat{Y}^{RB}$ is usually small, it can be much bigger for  $\mbox{mse}^{RB}$.
    \item Decreasing $n_1$ often decreases Monte-Carlo variance of the MSE estimator. This is because the consequent increase in $n_2$ improves the accuracy of the (conditional on $s_1$) design based MSE estimators, while it does not affect much the variance of $\mbox{mse}^{RB}$ as it is based on the full sample.
    \item The number of subsamples is usually fixed by the computational resources available, so it makes sense to tune $n_1$ to increase the efficiency of the MSE estimator. This might also affect the overall (not Monte-Carlo) performance of $\hat{Y}^{RB}$, so $n_1$ should not be too small.
\end{enumerate}

Let us see now the results from a simple example by \citet{ZSL24}. A population of size $N=1000$ was generated by $y_k = \beta_1 x_{1k} + \beta_2 x_{2k} + \epsilon_k$ with IID $x_{1k} \sim \mbox{LogN}(1,1)$, $x_{2k} \sim \mbox{Poisson}(5)$ and $\epsilon_k \sim N(0, \sigma^2/4)$, where $\sigma^2$ is the population variance of $x_{1k}$. Let $s$ be given by simple random sampling without replacement from this fixed population, where $n=100$. Let the mis-specified full-sample predictor be $\mu(x_1, s) = a + x_1 b$, where $(a,b)$ are the sample ordinary least square fit of $y$ on $x_1$ only.\\

\begin{table}[!htb]
\centering
\caption{MSE estimation from $250$ samples, $T=1000$, $\mu(x,s)$ for $\hat{Y}$ and $\bar{\mu}(x,s)$ for $\hat{Y}^{RB}$, (training, test) set of size $(n_1, n_2)$, RE against variance of HT-estimator.}
\label{tab:linMSE}
\begin{tabular}{p{1.5cm}p{2.0cm}p{1.6cm}p{2.0cm}p{1.6cm}p{2.0cm}}
\hline\noalign{\smallskip}
$(n_1, n_2)$ & $\mbox{MSE}(\hat{Y})$ & $\mbox{RE}(\hat{Y})$ & $\mbox{MSE}(\hat{Y}^{RB})$ & $\mbox{RE}(\hat{Y}^{RB})$ & $\mbox{CV}(\widetilde{\mbox{mse}}^{RB})$ \\
\noalign{\smallskip}\hline\noalign{\smallskip}
(98, 2)   & 386532.7 & 0.44 & 386632.4 & 0.44 & 3.48 \\
(80, 20) & 363613.9 & 0.41 & 363441.5 & 0.41 & 0.31 \\
(70, 30) & 362673.0 & 0.41 & 357146.9 & 0.41 & 0.21 \\ 
\noalign{\smallskip}\hline\noalign{\smallskip}
\end{tabular}
\end{table}

Table~\ref{tab:linMSE} shows the results of simulating MSE estimation based on 250 independent samples, given $T=10^3$, where the subsampling $q$-design is simple random sampling without replacement of $s_1$ from each realised sample given $n_2 = 2, 20, 30$. The MSE is simply the average squared error of either $\hat{Y}$ or $\hat{Y}^{RB}$ over the 250 samples, and the relative efficiency (RE) is the ratio between either MSE and the variance of the HT-estimator. Notice that the three $\mbox{MSE}(\hat{Y})$ here are all estimators of the same MSE, each using 250 independent samples, since $\hat{Y}$ depends only on $s$.\\

For $n_2$ up to 20 (or even 30), $\mbox{MSE}(\hat{Y}^{RB})$ is practically equal to $\mbox{MSE}(\hat{Y})$. The CV of the MC-MSE estimator $\widetilde{\mbox{mse}}^{RB}$ is drastically reduced by setting $n_2$ to 20 or 30 instead of 2. In comparison, the CV of the exact-RB MSE estimator $\mbox{mse}^{RB}$ is 0.14 by simulation, whereas the CV of the HT variance estimator is 0.32. This confirms that setting $n_2$ to be 20 (or even 30) and using a larger but practical $T$ would work satisfactorily for MSE estimation in this setup.\\

In terms of the choice of estimator, we notice that the mis-specified predictor $\mu(x_1, s) = a + x_1 b$ yields a design-based MSE that is less than half of the variance of the HT-estimator, and the bias of $\hat{Y}$ or $\hat{Y}^{RB}$ is in this case a negligible part of the MSE. Finally, as mentioned before, there is no reason why one cannot adopt $\bar{\mu}(x,s)$ in \eqref{RBest}, for which MSE estimation is unbiased, instead of using $\mu(x,s)$, as they show similar accuracy.\\

Some final remarks:
\begin{enumerate}
\item Design-based predictive inference from finite population probability sampling allows great flexibility in the choice of the estimator.
\item It provides design-based bias and MSE estimates that are valid independently from the assumptions of the predictor (if there were any).
\item So, actually, the model vs design controversy disappears. 
\item Providing a design-based MSE estimator looks like a natural extension to traditional variance estimator.
\item While here we focused on totals estimation, \citet{ZSL24}  also develop a design-based theory for estimations at the individual level.
\item It is a computationally demanding method, as it requires to reestimate the model multiple times.
\end{enumerate}

The highly predictive power of statistical learning models paves again the way to explore the combination of statistical models with design-based inference for estimation in finite populations. As stated by \citet{Tot24a}, model-assisted estimation techniques allow us to reduce the dependence on modelling assumptions, can lower variance over traditional linear models \citep{SSW92}, and, as shown above, when combined with subsampling, emulate train-test splits to estimate and correct errors.

\subsection{Selective and macro editing}
\label{sub:SelMacEdit}
Statistical data editing is the production task devoted to detect and treat errors in order to gain quality for the estimation phase \citep{deWPanSch11a}. It may consume up to 20\,\% of the production resources \citep[see e.g.][]{JonLoo18a}. The complexity of the design, development, and execution of so-called editing and imputation (E\&I) strategies has been growing since the early years of clerical work \citep[see e.g.][]{UNE94, UNE97, EDI07a, UNE19a}. High-level editing business functions, i.e. editing modalities are nowadays assembled to deal with the different types of error present in different statistics.\\

In this resource-intensive scenario, more efficient forms of data editing are continuously searched, and selective editing stands as a relevant choice for efficiency gains \citep{PanSchLoo13a}. Selective editing \citep{deW13a} allows the statistician to concentrate high-valued resources on errors having a substantial influence on final estimates, thus increasing quality while being cost-efficient. The basics of this editing modality amounts to assigning a local (item) score $s_{k}^{(q)}$  to each variable $Y^{(q)}$, $q=1,\dots,Q$, under inspection, combining these into a global (unit) score $S_{k}=S(s_{k}^{(1)},\dots, s_{k}^{(Q)})$, computing a threshold $t$, and, finally, selecting units above the threshold $t$  for interactive editing, i.e.\ selecting $\{k\in s\subset U: S_{k}\geq t\}$.\\

By and large, no accepted theory for selective editing exists. In fact, selective editing is an
umbrella term for several methods to identify the errors that have a strong influence on final estimates. In our optimization approach to selective editing \citep{ArbRevSal13a}, local scores can be understood as conditional expectation values of measurement error models:

\begin{equation}\label{eq:ItemScore}
	s_{k}=d_{k}\cdot\mathbb{E}_{m}\left[|Y_{k}^{\textrm{raw}}-Y_{k}^{0}|\big| \mathbf{Z}^{\textrm{aux}}\right],
\end{equation}
\noindent where $d_{k}$ denotes the design weight of unit $k$, $Y_{k}^{\textrm{raw}}$ stands for the raw (unedited) value of target variable $Y$ and $Y_{k}^{0}$ for its true value, $\mathbf{Z}^{\textrm{aux}}$ denotes the available auxiliary information, and $m$ stands for the underlying assumed measurement error model.\\

This expression opens the door to using supervised statistical learning models provided a historic dataset with both raw and validated values for the same variable is available \citep[see][for a different approach]{ForGar25a}. Furthermore, the versatility of these models allows us to compute local score values both for categorical  and continuous  variables. In the former case the identity \eqref{eq:ItemScore} reduces to 

\begin{equation}\label{eq:Score_Categ}
 s_{k}= d_{k}\cdot\mathbb{P}\left(Y_{k}^{\textrm{raw}}\neq Y_{k}^{0}\big|\mathbf{Z}^{\textrm{aux}}\right),
\end{equation}

\noindent where $\mathbb{P}(Y_{k}^{\textrm{raw}}\neq Y_{k}^{0}\big|\mathbf{Z}^{\textrm{aux}})$ denotes the error probability for unit $k$ in the variable under inspection. Thus, the problem reduces to a probability estimation problem, which we have approached using random forests. This choice is due to its good properties (versatility, performance, robustness, ...), see e.g. \citet{BarDhiTimBouCal24a}. Ultimately, model selection should be carried out investigating different choices (logistic regression, boosting, etc.). We have used this approach in the editing phase of the Spanish branch of the European Health Interview Survey for the so-called `social class` variable, which is indeed a subject-matter aggregation of the ISCO occupation code. These error probabilities are estimated using a random forest on the set of variables used during the clerical revision of the questionnaires (age, sex, economic activity, educational degree, professional situation, job situation, ...). The target variable in the random forest model is the measurement error indicator $I(y_{k}^{\textrm{raw}}\neq y_{k}^{0})$, where the training data is taken from the historic dataset of raw and validated values in the survey. In actual production conditions, once a record is clerically revised, thus producing both a new pair of raw and validated values, the record enters into the training data set for a new random forest updating to be applied to the new data collection batch. The (unrevised) questionnaires are then given an editing priority upon their collection every time a new data collection batch is processed, so that clerical revision in the interactive editing modality is rationalised.\\

\begin{figure}[!htb]
\centering
\includegraphics[width=0.43\textwidth]{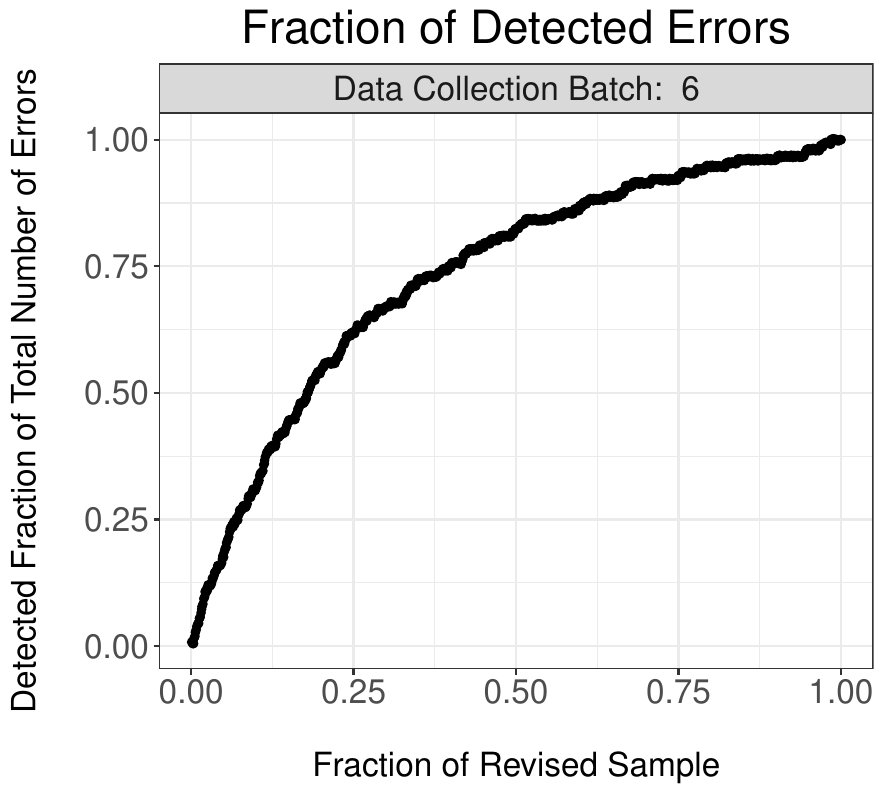}
\includegraphics[width=0.43\textwidth]{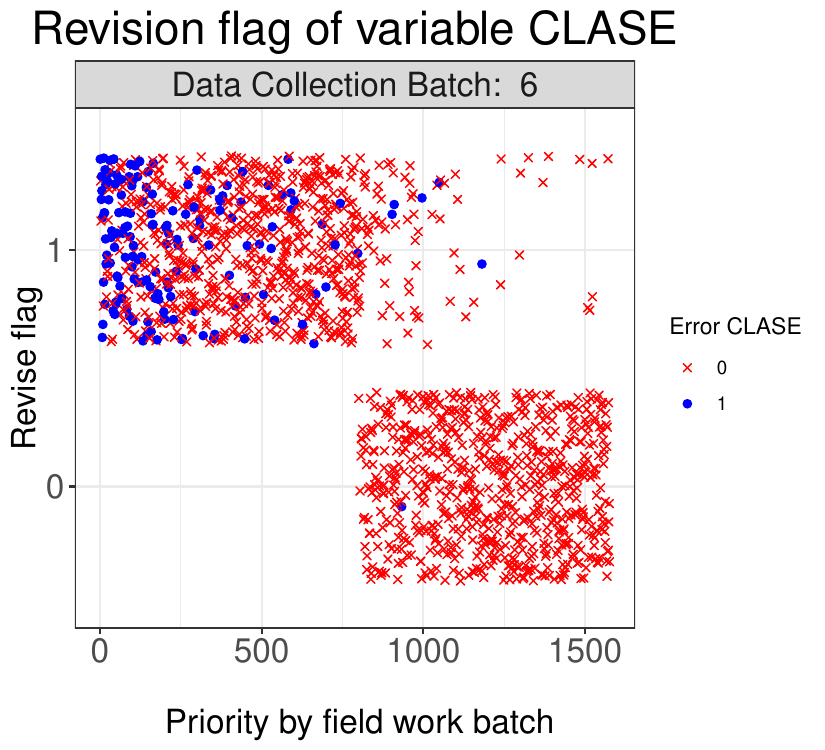}
\caption{(Left) Fraction of detected measurement errors in the categorical variable \texttt{CLASE} vs.\ fraction of revised units according to the prioritization provided by the local score (data collection batch no.\ 6). (Right) Revision flags ($0$ or $1$) vs. questionnaire (editing) priority from the score value $s_{k}$. Shape (cross or solid circle) indicates the absence or presence of error in each unit of data collection batch no.\ 6.} 
\label{fig:RevFlagEHIS} 
\end{figure}

In Figure~\ref{fig:RevFlagEHIS} we include a visualization of the prioritization performance by the local score $s_{k}$ in Equation~\eqref{eq:Score_Categ} for an arbitrary data collection batch. In the left figure we represent the fraction of detected errors as we revise questionnaires in the descending order derived from $s_{k}$ in terms of the fraction of sampling units to revise. In the right figure we represent those units marked for clerical revision (revision flag $= 1$) together with the absence/presence of errors upon inspection. The threshold was established by the subject matter experts after the first 5,000 sampling units were systematically revised in this pilot experience based on the experimental finding that around the first 50\,\% of the prioritised sample contained around 75\,\% of the total number of erroneous units. The editing tasks were later completed during the macro editing phase.\\

This approach can also be adapted for continuous variables. Suppose the target variable $Y$ represents the turnover of a statistical unit. Then, the measurement error $\epsilon_{k}=y_{k}^{\textrm{raw}}-y_{k}^{0}$ is indeed a semi-continuous variable, since it takes either the value $0$ or a continuous value. Indeed we can introduce (i) a binary variable $\delta_{k}^{(\epsilon)}$ taking the value $1$ when $\epsilon_{k}\neq 0$ and the value $0$ otherwise and (ii) a continuous variable $e_{k}^{(\epsilon)}$ with the non-null measurement error value when $\delta_{k}^{(\epsilon)}=1$. Thus, we can write $\epsilon_{k}=\delta_{k}^{(\epsilon)}\cdot e_{k}^{(\epsilon)}$, and
starting from Equation \eqref{eq:ItemScore} we can decompose 

\begin{eqnarray}
s_{k}&=&d_{k}\cdot\mathbb{E}_{m}\left[\delta_{k}^{(\epsilon)}\cdot |e_{k}^{(\epsilon)}|\ \big| \mathbf{Z}^{\textrm{aux}}\right]\nonumber\\
&=&d_{k}\cdot\mathbb{E}_{\delta}\left[\mathbb{E}_{e}\left[\delta_{k}^{(\epsilon)}\cdot |e_{k}^{(\epsilon)}|\ \big|\delta_{k}^{(\epsilon)},\mathbf{Z}^{\textrm{aux}}\right]\right]\nonumber\\
&=& d_{k}\cdot\mathbb{P}\left(\delta_{k}^{(\epsilon)}=1\big| \mathbf{Z}^{\textrm{aux}}\right)\cdot \mathbb{E}_{e}\left[|e_{k}^{(\epsilon)}|\ \big|\delta_{k}^{(\epsilon)}= 1,\mathbf{Z}^{\textrm{aux}}\right].
\end{eqnarray}

We can then proceed in two steps. Again assuming a historic dataset with both raw and validated values of this target variable is available for the same survey, firstly, we can estimate the error probability $\mathbb{P}(\delta_{k}^{(\epsilon)}=1\big| \mathbf{Z}^{\textrm{aux}})$. Then, in a second step, using those units with measurement error as training data, we can adjust a second regression model to predict $|e_{k}^{(\epsilon)}|$. As features for these models we use both classification and other target survey variables and paradata as well as new variables derived thereof \citep{Boh20a}, i.e.\ all available information. The guiding principle is to consider the variables taking into account a golden-standard editing procedure.\\

In addition, the value $y_{k}^{\textrm{raw}}$ can also be missing, which obliges us to propose a specific regression model for the measurement error $|\epsilon_{k}|$ of these units. Once the score values $s_{k}$ are available for the entire sample, an editing priority in terms of their descending values can be assigned to each questionnaire.\\

\begin{figure}[!htb]
\centering
\includegraphics[width=0.75\textwidth]{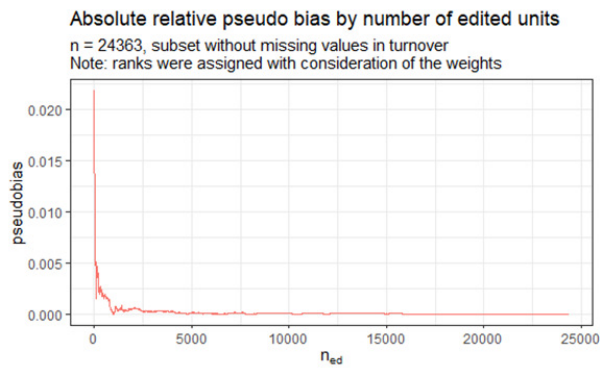}
\includegraphics[width=0.75\textwidth]{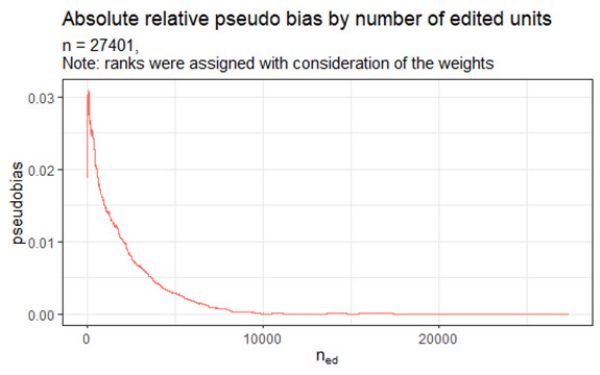}
\caption{\label{fig:IASS} (Top) Reduction of the relative pseudo-bias in absolute value in terms of the number of revised units. Units with missing values are not included in the list of units to revise. (Bottom) Reduction of the relative pseudo-bias in absolute value in terms of the number of revised units. Units with missing values are included in the list of units to revise. Results correspond to the consultations test set.} 
\end{figure}

In Figure~\ref{fig:IASS} we represent the reduction of the relative pseudo-bias in absolute value $ABS =\frac{\big|\hat{Y}(n_{(ed)})-\hat{Y}^{(val)}\big|}{\hat{Y}^{(val)}}$, where $\hat{Y}(n_{(ed)})=\sum_{k\in s_{ed}}d_{k}y_{k}$ and $s_{ed}$ denotes the sample of units with $n_{ed}$ revised units and $n-n_{ed}$ units with raw values and $\hat{Y}^{(val)} = \hat{Y}(n)$, in terms of the number of revised units in the priority order derived from the score values. These are the results of a pilot study with real data from the Spanish short-term business statistics called Service Sector Activity Indicators Survey \citep{Boh20a}. You can observe that efficiency gains can be obtained by focusing on those units with large score values.\\

Some remarks:

\begin{itemize}
    \item Historic datasets with raw and validated values are a rich source of information to predict measurement errors to gain efficiency during the editing phase. Both the data architecture and the process design and execution in a statistical office should be oriented towards their preservation and storage. Statistical learning models constitute a promising tool to dive into the data error patterns present in the process.
    \item Although hyperparameter optimization and model selection are compulsory steps in statistical learning model adjustment, the ultimate goal in the production process should drive the modelling exercise. In our case, random forests have provided enough quality to get some first efficiency gains using editing quality indicators as figures of merit (error detection rate, pseudo-bias, ...).
    \item This proposal covers only the computation of local scores. Their combination to obtain global scores can be carried out as in the traditional approach. However, a more natural generalization can be provided in the following way. For categorical variables, instead of estimating $\mathbb{P}(Y_{k}^{\textrm{raw}}\neq Y_{k}^{0}\big|\mathbf{Z}^{\textrm{aux}})$ for a single variable $Y$, we can estimate $\mathbb{P}(Y_{k}^{(1)\textrm{raw}}\neq Y_{k}^{(1),0} \lor \dots \lor Y_{k}^{(Q)\textrm{raw}}\neq Y_{k}^{(Q),0}\big|\mathbf{Z}^{\textrm{aux}})$ for $q=1,\dots, Q$ variables, so that error detection in any of these variables is accomplished. For continuous variables further research is needed.
    \item An indicator for the efficiency of the prioritization in order to compare different methods is needed (e.g. related to the area under the curves in Figure~\ref{fig:IASS} \citep[see][]{ArbRevSal13a}).
    \item Not only do efficiency gains impinge on the cost-efficiency dimension of production quality but also it allows the process to be more timely. We claim that the predictive power of Machine Learning techniques could in principle allow us to improve some editing business functions to the point that the editing process could a priori be redesigned to consider the production and release of some fast early estimates with influential errors under control (see also Section~\ref{sub:EarlyImput} below). Considering the Generic Statistical Data Editing Model (GSDEM) v2.0 \citep{UNE19a} as the natural framework to design and develop editing and imputation strategies, any method involving a prediction will be improved by using these statistical learning models. Nonetheless, microdata quality in their final validated form is crucial both for final estimation of the population quantity of interest (population total in this case), for training this kind of models, and for researchers and stakeholders accessing these data.  
\end{itemize}

\subsection{Statistical classification coding}\label{fd-ref::scc}
\label{sub:ClaCod}

A statistical classification is defined as a \textquotedblleft hierarchically organised set of mutually exclusive and jointly exhaustive categories that share the same or similar characteristics, used for meaningfully grouping the objects or units in the population of interest\textquotedblright\ \citep{GSIMv20}. It constitutes an extremely valuable statistical instrument extensively used by statistical offices as well as by other public organizations and even stakeholders both for statistical production and analysis.\\

They are commonly arranged in classification series, which in turn are grouped into classification families based on common concepts \citep{GSIMv20}. The family of economic activity classifications probably conforms the most widely used example of statistical classification in the production of official statistics. We shall focus on this family in the Spanish context, which runs completely similar to many other countries.\footnote{See e.g. \citet{fd-chapter-atdenace}, \citet{fd-chapter-insee} and \citet{fd-chapter-ba} in this book.} The global reference is provided by the International Standard Industrial Classification of all economic activities (ISIC) \citep{ISICv50}, which is adapted to European needs in the \emph{Nomenclature statistique des activités économiques dans la Communauté Européenne} (NACE) \citep{NACEv21}, and it is further specialized for the Spanish context in the \emph{Clasificación Nacional de Actividades Económicas} (CNAE) \citep{CNAE2025}. This will be the focus of this section.\\
 
ISIC, NACE and CNAE have undergone a thorough revision during the past years. This process has produced the ISIC Rev.~5, the NACE Rev.~2.1 and the CNAE-2025. A revision process of a statistical classification strongly impinges on many aspects of production and dissemination of official statistics. The core of this challenge is the need to assign economic activity codes in the new versions of the classification for all statistical units, hence also to aggregates and indices, in all produced and released statistics. Furthermore, in the Spanish context, the CNAE-2025 is intensively used for non-statistical purposes by other public bodies, especially the National Tax Agency and the Ministry of Social Security. Researchers and analysts also see themselves pushed to make use of the new classification. This fact, in turn, implies that citizens, researchers, and firms are compelled users of the classification.\\

The increasing use of administrative data for the production of official statistics, data thoroughly generated by administrative bodies, makes it advisable for statistical offices to focus on the quality of the generation process of economic activity codes. In this context, Statistics Spain has undergone two complementary strategic actions. On the one hand, the use of CNAE-2025 has been made legally compulsory to classify and code any economic activity variable in all public administrative registers \citep{CNAE2025BOE}. This will ease the reuse of this data for official statistical purposes and, at the same time, will boost semantic interoperability in the Spanish Public Administration. On the other hand, Statistics Spain has developed and is maintaining and evolving an automatic coding tool named \textrm{CodIA} assisting in the selection of a CNAE-2025 activity code from a textual description of an economic activity \citep{CodIA}.\\ 

\textrm{CodIA} has been built using Natural Language Processing (NLP) techniques. In the following we shall describe the main aspects of technical and strategic interest of this process. Firstly, since the structure of statistical classifications is similar in different classification series (e.g.\ occupation, education, ...), the coding tool should abstract the semantic content of the classification and benefit from its metainformation elements such as explanatory notes, introductory guidelines, and similar. In this sense, \textrm{CodIA} has been separately trained and tested using data regarding both the CNAE-2009 and the CNAE-2025 \emph{mutatis mutandi}. In this sense, we shall always refer to CNAE embracing both versions, unless otherwise specified. This mindset allows us to face the challenge of building a tool which can be easily replicated for other statistical classifications and future versions of CNAE.\\

 Secondly, the hierarchical structure in sections, divisions, groups, and classes can be approached either top-down or bottom-up. In the former, the task is to find the code for the section (1-digit code) and move downwards providing then the code for the division (2-digit code), for the group (3-digit code), and finally for the class (4-digit code). In the latter, the task is to find the code for class, and making use of the nested hierarchy, to automatically provide the corresponding codes for the section, group, and class. After very short-ranged and preliminary tests suggesting so, for the interest of time, we decided to follow the bottom-up approach: \textrm{CodIA} provides CNAE codes for the class category of a given textual description of an economic activity\footnote{See \citet{BerWydCheBat24a} for results of a thorough comparison of both approaches in the context of the occupation classification family.}. These codes are provided with corresponding probability-like scores $s_{c}(text)$ for each returned class code $c$ and an input text $text$. For nesting categories such as section, division, and group, the corresponding scores are elementarily computed by aggregation: $s_{g}(text)=\sum_{c\in g}s_{c}(text)$, where $g$ denotes a group, and similarly for divisions and sections. Thus, results hereafter will focus on the class level, unless otherwise explicitly mentioned. More specifically we shall focus on the CNAE-2025.\\
 
 Thirdly, in the need to recode statistical units under the revised version of the classification, in the case of CNAE-2025 \textrm{CodIA} also accepts optionally the class activity code of CNAE-2009 as an input together with the textual description, thus assisting in recoding tasks.\\

At first sight one might think that automatic coding is an already solved problem as a text classification application \citep[see e.g.][]{KamLiuWhi19a}. We just need curated training sets and a state-of-the-art NLP model. One may even think that a large language model (LLM) may do it for us: \textquotedblright just ask ChatGPT\textquotedblright. However, automatic coding of statistical classifications in the context of official statistical production presents some non-trivial issues:
\begin{itemize}
	\item The task of classifying and coding descriptions of economic activities is hard even for a human expert for the following reasons:
	\begin{itemize}
		\item The number of categories (classes) is very high (664 in the CNAE-2025).
		\item The boundaries between classes are complex, subtle, and sometimes blurry.
		\item Many descriptions are incomplete or ambiguous.
	\end{itemize}
	\item Single classes (lowest hierchical categories) contain indeed an unequal variety of economic activities. 
	\item The coding tool must produce satisfactory results for different types of inputs depending  on the needs of the user and/or the source of the textual description to be coded. In real production conditions we may find decriptions with a single word or concept, while others may provide a long detailed paragraph.
\end{itemize}
	
Thus, the first decision is to choose a family of NLP models. To this end, we carried out a comparison between the most popular options. A summary of the output of this analysis is presented in Table~\ref{tab:NLPmod}.\\

\begin{table}[!htb]
    \centering
	\caption{Comparison of family of NLP models for automatic classification coding}
	\label{tab:NLPmod}
	\begin{tabular}{p{2.5cm}p{2mm}p{4.4cm}p{2mm}p{4.4cm}}
		\hline\noalign{\smallskip}
		\textbf{Family of Models} & & \textbf{Main Advantages} && \textbf{Main Disadvantages} \\
		\noalign{\smallskip}\hline\noalign{\smallskip}
		Bag-of-words + “classic” ML model   &  & - Easy to implement\hfill\par
			- No confidentiality issues
		 && - Performance far below state-of-the-art\\\hline
		Fasttext \citep{JouGraBojMik17a}   &  & - Very easy to implement\hfill\par
		- Very short training time\hfill\par
		- No confidentiality issues && - Performance below state-of-the-art\\\hline
		BERT-like \citep{LiuOttGoyDuJosCheLevLewZetSto19a}   &  & -	Almost state-of-the-art performance && 	- Training and deployment are time and resources demanding\hfill\par
		- Careful metaoptimization is needed\\\hline
		Zero-shot LLM  &  & - Very easy to implement && 	-  Performance far below state-of-the-art\hfill\par
		- Confidentiality issues\\\hline
		LLM + RAG  &  & - Probably state-of-the-art performance && 	- Demanding in human effort, time and computational resources\\\hline		
		LLM + Fine tuning  &  & - State of the art performance && 	- Training and deployment are time and hugely resources demanding\\\hline
		
		\noalign{\smallskip}\hline\noalign{\smallskip}
	\end{tabular}
\end{table}

With this analysis in mind, taking into account (i) the scarcity of available computational resources, (ii) the time needed to collect and prepare enough training and test datasets, (iii) the deadline to deploy the solution, (iv) the robustness advised for technological solutions maintained in production conditions given the available resources, and (v) the inspiring experience and counselling by the French National Statistical Institute\footnote{We acknowledge the invaluable contact with Romain Lesur and INSEE's Data Science Lab. See  \url{https://ssplab.lab.sspcloud.fr.}} \citep{FarSei23a}, we decided to use Fasttext. An additional consideration should be made here: BERT-like models represent a huge advancement over Fasttext, or other simpler and older models, because they incorporate the attention modules. However, we hypothesize that this advantage is reduced in our context because most texts are very short in the 1-10 words range.\\

Once the family of models was chosen, the next step was to build datasets for training and validation. By and large, pairs of text-code instances needed to be collected. Notice that differences must be taken into account between the CNAE-2009 and the CNAE-2025, because the latter is new and no data was available. For the CNAE-2009 data were collected from collection and editing paradata of existing official business statistics in Statistics Spain. Both structural and short-term business statistics were used. For the CNAE-2025, a two-fold course of action was undertaken. On the one hand, taking advantage of the existing infrastructure for data collection, two large-sample structural business surveys were extended to include a specific questionnaire item requesting business units to provide their activity code according to the CNAE-2025. On the other hand, a specific ad-hoc survey was implemented with a sample size of 110,000 business units with a 1:N correspondence between their CNAE-2009 and CNAE-2025 main activity codes. In this first step, in total, datasets amounting to around 420,000 instances were obtained. A test set with around 1,000 instances was carefully revised by experts in order to ensure maximum coding quality. As a consequence, useless or incomplete descriptions were discarded.\\

At this point, textual descriptions were pre-processed; Fasttext metaparameters were optimized (validating the different combinations with the validation set) and a first Fasttext model was trained with the whole training and validation dataset. As an automatic coding tool providing just one single code for each textual description, \textrm{CodIA} reached a global accuracy of $0.63$. A careful analysis of this performance, both quantitative and qualitative, showed that, while the overall performance was not poorly bad so that this model could possibly work for manual coding assistance, it presented some weak points:
\begin{itemize}
	\item The performance over minority classes was very poor. This did not heavily affect the overall metrics and use of the model because of the actual occurrence of these classes in the  Spanish economy. However, it could erode the credibility of the coding tool.
	\item This model was unable to detect many concepts clearly related to one or few classes because they did not occur or occur very seldomly in the training set. Some of these concepts are, however, present in the titles and/or explanatory notes of the CNAE.
	
\end{itemize}

As a second step, thus, we decided to generate synthetic data to enrich the training dataset by applying three strategies:

\begin{itemize}
	\item We added the titles and explanatory notes themselves, paired with the corresponding codes, as instances.
	\item We used LLMs to build a dictionary with word synonyms and we replicated the original textual descriptions by replacing some of its words by these synonyms.
	\item We asked LLMs via API to create synthetic economic activity descriptions. For this, we carefully tuned the prompt, including the title and explanatory notes of a given class as well as detailed instructions and examples to optimize the behavior of the LLM.
\end{itemize}

Additionally, we trained the model initialized with pre-trained weights for the Spanish language, with the ambition that it would grasp semantic relations, including those not occurring in the dataset.\\

With all these initiatives, we got a dataset of about $2.5$ million instances. After training the Fasttext model with the previous metaparameters over this larger dataset, the global accuracy as an automatic coder providing a single code was raised up to $0.69$. A qualitative analysis showed that many of the problems listed above were softened, even despite the huge imbalance of the training dataset. The performance of the minority classes was still worse than the majority ones and some unusual concepts were not treated correctly. However, it is worth noting that this model could be trained in a laptop in less than $10$ minutes, which is very convenient for production conditions and model updating and retraining.\\

Finally, before deploying the tool into production, we undertook a stricter testing procedure. To this end, we collected user consultations received through Statistics Spain's corporate mailbox asking which code corresponded to a given provided description. It should be noted that these descriptions are much harder to classify and code than the average. These consultations, which are solved by classifications experts, are indeed motivated by this difficulty. The global accuracy over this combined dataset resulted in $0.53$, while we got $0.41$ when using the model trained only with real data. We could observe that the improvement of adding synthetic data is more acute in this case. We believe that the use of synthetic data for training language models is a strategic approach which should be further pursued.\\

At this point \textrm{CodIA} can be used as an automatic coder or as a coding assistant. As an automatic coder it is configured to return the class code with the highest probability-like score. As a coding assistant it is configured to return the class codes scoring above a chosen threshold. Notice that in either case there will be textual descriptions possibly returning no class code because the model is configured to return class codes only if they are scoring above the threshold. With the choice of the threshold the user can impose more or less stringent conditions on venturesome outputs.\\

To illustrate the performance of \textrm{CodIA} we focus on four figures of merit:

\begin{itemize}
	\item Precision, understood as the fraction of input instances resulting in the correct class code (automatic coder) or in a set of class codes containing the correct class code (coding assistant).
	\item Recall, understood as the fraction of input instances resulting in at least one class code. If recall equals $1$, \textrm{CodIA} always returns at least one output class code.
	\item Threshold, selected by the classification expert to provide the minimum value of the probability-like scores of input instances producing an output class code. As an automatic coder, \textrm{CodIA} will produce either $1$ or $0$ output class code. As a coding assistant, it will produce $0$ or more output class codes.
	\item Averaged number of output class codes, understood as the mean of the number of output class codes produced for each input instance. Notice that this figure of merit only makes sense when \textrm{CodIA} is used as a coding assistant.
\end{itemize}

\begin{figure}[!htb]
	\includegraphics[width=\textwidth]{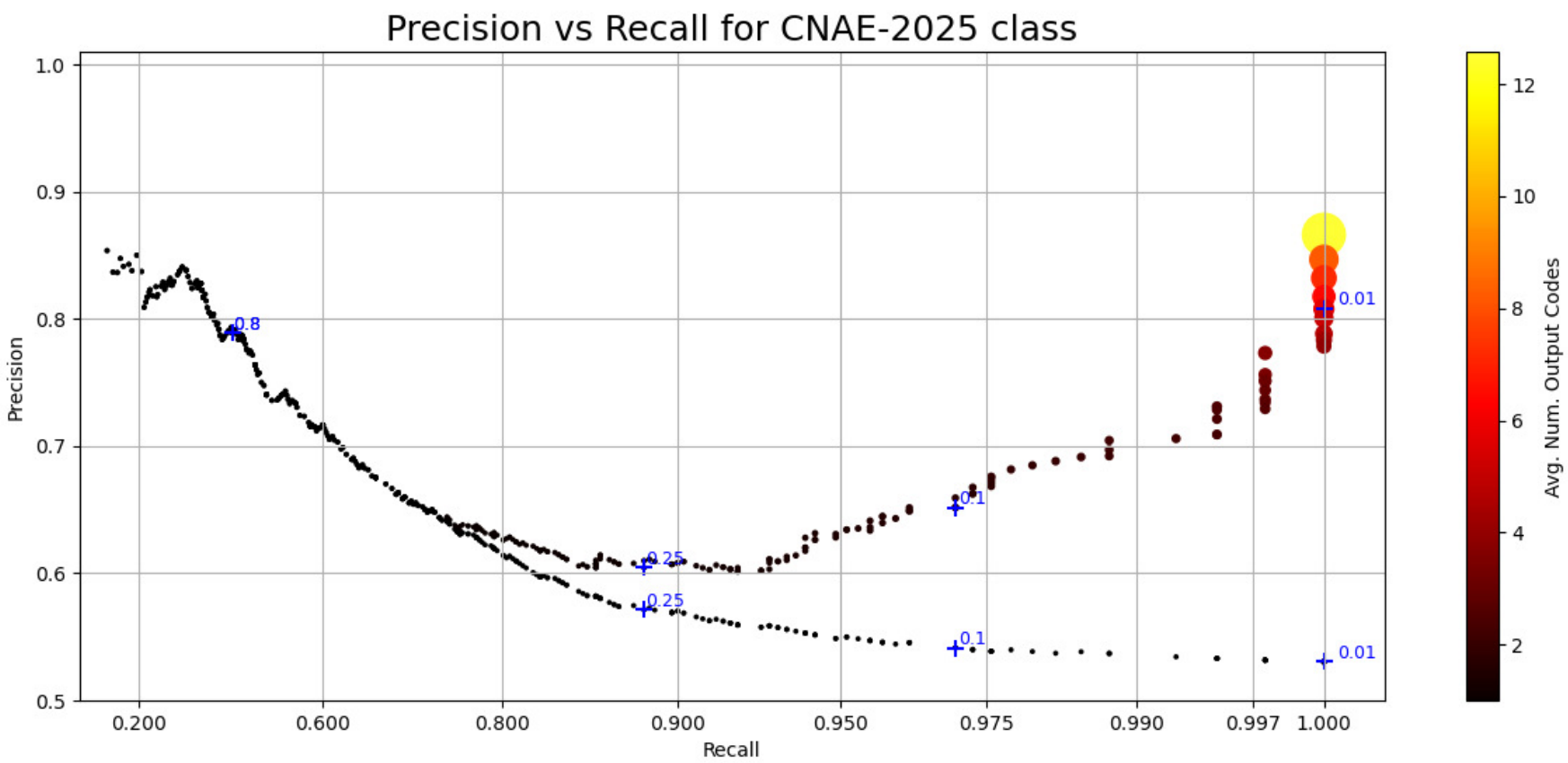}\\
	\includegraphics[width=\textwidth]{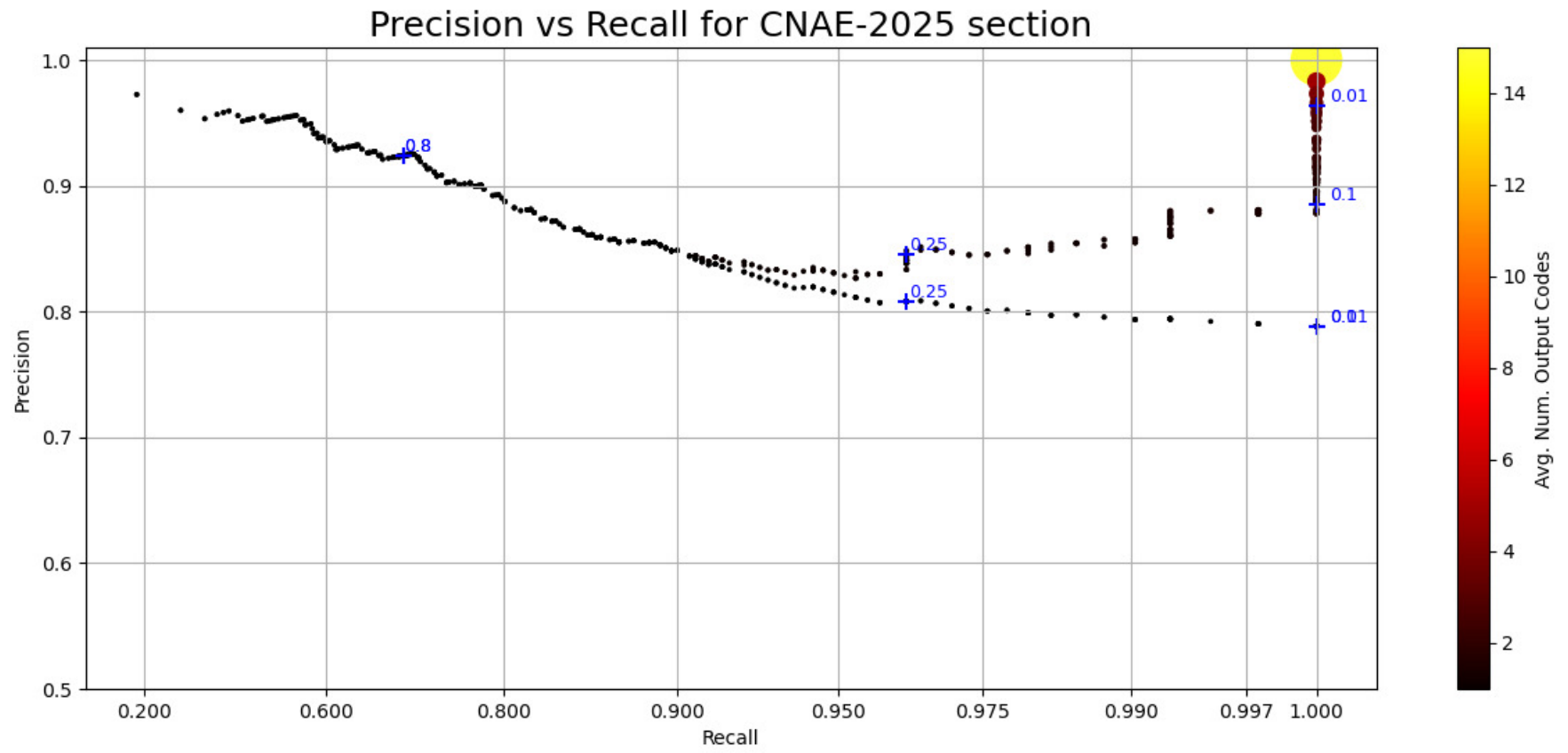}
	\caption{Results of \textrm{CodIA} as an automatic coder and as a coding assistant (see text) for CNAE-2025 class codes (above) and section codes (below). Notice the logarithmic-like scale in the x-axis.} 
	\label{fig:codIA} 
\end{figure}

The results on the combined dataset with real and synthetic data are illustrated in Figure~\ref{fig:codIA}. The graph represents the performance of \textrm{CodIA} both as an automatic coder (lower curve) and as a coding assistant (upper curve). From left to right results correspond to decreasing values of the threshold. Four threshold values are explicitly marked. Notice how \textrm{CodIA} diminishes its performance when forced to provide always a single output class code (lower curve). Notice also how it starts returning more than one output class code when the threshold is chosen below a given value. Indeed, as a coding assistant, the lower the threshold, the higher the number of output class codes. You can observe that with a very low threshold, it always returns output class codes (recall equals $1$) providing $8$ output class codes and $14$ output section codes on average, out of which around 85\,\% and 94\,\%, respectively, contain the correct answer.\\ 

\begin{figure}[!htb]
	\includegraphics[width=\textwidth]{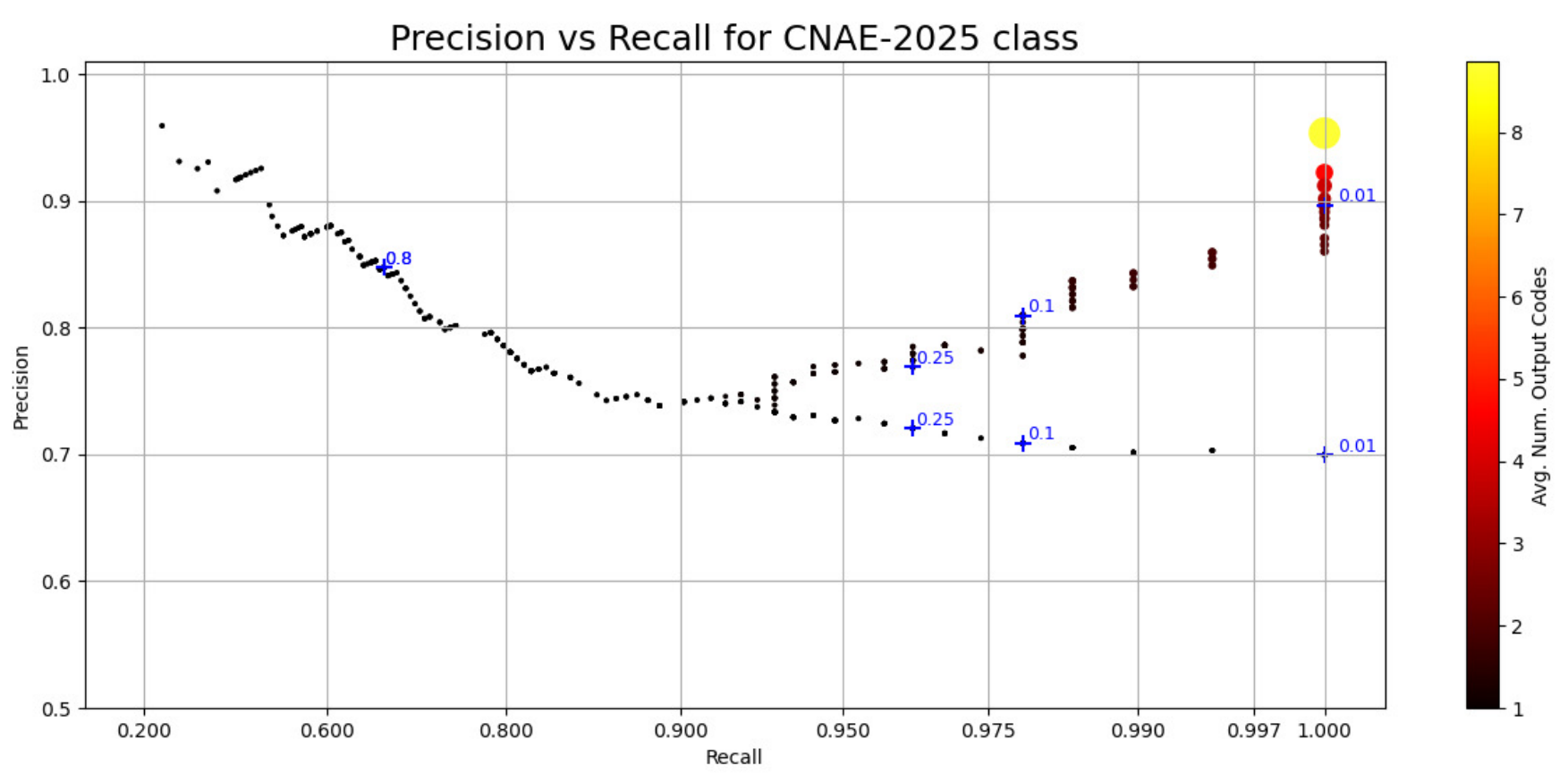}
	\caption{Results of \textrm{codIA} as an automatic coder and as a coding assistant in production for CNAE-2025 class codes. Notice the logarithmic-like scale in the x-axis.} 
	\label{fig:codIAprod} 
\end{figure}

With these results, \textrm{CodIA} was deployed into production \citep{CNAE2025} on January 15th, 2025. In the first $20$ days more than 20,000 requests were processed in Statistics Spain's website. A sample of them was manually labelled and the preceding four figures of merit were recomputed. Results are illustrated in Figure~\ref{fig:codIAprod}.\\

To sum up, NLP techniques constitute a promising breadth of tools to handle statistical classifications, already producing results to improve classifying and coding activities. More research and analyses need to be carried out to disentangle and understand challenges as the construction of high-quality training sets or the blurry limits between textual descriptions  of some economic activities with different categories. Although the choice of language models is important, we detect a need to clearly quantify and unravel the quality of this type of data for statistical purposes.

\section{New business functions for more granular, frequent, and timely statistics}
\label{sec:NewBusFunc}
Traditional demands on official statistics, underlined in crisis periods as e.g. in the last financial crisis and COVID pandemic, is the production and release of more granular, more frequent, and more timely statistics \citep{Hol07a, CES23a}. These demands are usually at the root of the proposals to use new data sources, especially big data and any other sort of digital data. This situation is especially visible and critical when producing survey statistics. The production of survey statistics requires in GSBPM's terminology to periodically collect, to process, to analyse, and to disseminate microdata and aggregates, where lower-level business functions are to be executed. To improve granularity directly we would require larger sample sizes with a prohibitive cost. To increase frequency we would need microdata to be more time-disaggregated (sometimes not even possible because of the definition of the target variables themselves). To improve timeliness we would need to execute these business functions at an extraordinary speed (e.g.\ could we request data providers to deliver their response to statistical offices in the first week of each month?).\\

In this section we present proposals and proofs of concepts to improve these quality dimensions using survey data or a combination of survey and administrative data. The core of the proposals is the use of statistical learning models with survey and administrative data (already present and available in most statistical offices) to overcome the slow delivery of statistics based on them. This paves the way for their integration with new digital data sources, a priori more frequent and more granular.

\subsection{Early imputation}\label{spain:sec:earlyImp}
\label{sub:EarlyImput}
Short-term business statistics constitute a key set of indicators produced by statistical offices to monitor different aspects of the economic situation in a country. They are subjected to a high pressure to inform about the inmediate state of the economy. When produced using survey data, the statistical process, broadly spreaking, needs to execute data collection, data editing, estimation and index computation, and dissemination and communication. For monthly statistics, data collection takes 3 to 4 weeks at least, to be followed by editing and the computation of indices.\\

In this line of thought, we propose to make use of statistical patterns in the historical microdata to predictively construct every monthly sample along with the execution of the data collection and data editing processes so that we have a synthetic microdata (survey + predicted) set at every time together with an assessment of the uncertainty. In this way, we can compute an early estimate of all those indices to be released later when the whole production process finishes.\\

We apply this proposal to the Spanish Industrial Turnover Index, subjected to European Regulation in the European Statistical System \citep{ICN_Reg}. Let $z_{k}^{my}$ denote the value of the turnover of an industrial establishment $k\in s^{my}$ for the reference time period with month $m$ and year $y$ from sample $s^{my}$. Firstly, we need to introduce the time dependence $z_{k}^{my}(\tau)$ within the data collection period measured in number of days so that $\tau=d$ stands for the date $my + d\textrm{ days}$ ($d$ days after the reference time period is over). We notice the following:

\begin{enumerate}
    \item The value $z_{k}^{my}(\tau)$ evolves along with the data collection and data editing from a missing value to a validated value going through different editing status (raw, edited during collection, edited interactively, etc.).
    \item If we denote the press release date by $d_{\textrm{release}}$, when $\tau\geq d_{\textrm{release}}$, it turns out that we have nonresponse or the value could not have been possibly validated or the value is finally validated. In the first two cases, imputation is applied to obtain a synthetic value $\hat{z}_{k}^{my}(\tau)$ under a given imputation model.
    \item We shall denote by $s^{my}(\tau)$ the set of sampling units having provided response up to time $\tau$ (response set) and $\hat{s}^{my}(\tau)=s^{my}-s^{my}(\tau)$.
    \item In practice, three batches are made available to subject matter experts and processed by them during data collection at $\tau=20, 29, 38$. The press release is published at $\tau_\textrm{release} = d_{\textrm{release}} = 51$ on average.
    \item Occasionally, responses arrive after the release date. These data are also processed and validated for revised versions of the indices. We shall label values with this final state of validation by $\tau=\tau_{f}$.
    \item No consideration about the sample selection or the index computation has been done so far. We are focusing on the microdata level for the time being.
\end{enumerate}

The core idea is to consider, for each reference time period $my$ in turn, all past monthly validated microdata sets $\mathcal{F}^{my}_{\leftarrow}=\{z_{k}^{\bar{m}\bar{y}}(\tau_{f})\}_{k\in s^{my}}^{\bar{m}\bar{y}<my}$, the collected and partially edited values $\mathcal{F}_{\updownarrow}^{my}(\tau)=\{z_{k}^{my}(\tau)\}_{k\in s^{my}(\tau)}$ at $\tau=20, 29, 38$ and territorial and sectorial classification microdata and collection paradata of each statistical unit $\mathcal{I}_{\leftarrow}^{my}=\{\mathbf{x}_{k}^{\bar{m}\bar{y}}\}_{k\in s^{my}}^{\bar{m}\bar{y}< my}$ and $\mathcal{I}_{\updownarrow}^{my}=\{\mathbf{x}_{k}^{my}\}_{k\in s^{my}(\tau)}$. These include continuous, semicontinuous, and categorical variables.\\

Next, we define $\mathcal{F}^{my}(\tau)=\mathcal{F}_{\leftarrow}^{my}\bigcup\mathcal{F}_{\updownarrow}^{my}(\tau)\bigcup\mathcal{I}_{\leftarrow}^{my}\bigcup\mathcal{I}^{my}_{\updownarrow}$. Notice that values from the past history are already validated (possibly not in their final form, but mostly) in preceding executions of the production process and values from the current time period correspond to responding units. Thus, no missing values are present in the dataset.\\

For each unit $k\in s^{my}$ and groups of units we carry out different computations to obtain three kinds of derived variables so that the set of features $\bar{\mathcal{F}}^{my}(\tau)$ can be decomposed as $\bar{\mathcal{F}}^{my}(\tau)=\bar{\mathcal{F}}^{my}_{\leftarrow, \textrm{unit}}\bigcup\bar{\mathcal{F}}^{my}_{\leftarrow, \textrm{aggr}}\bigcup\bar{\mathcal{F}}^{my}_{\updownarrow, \textrm{aggr}}(\tau)$, where:

\begin{itemize}
    \item  $\bar{\mathcal{F}}^{my}_{\leftarrow, \textrm{unit}}$ contains features for each unit $k\in s^{my}$ constructed with variables of the same unit. For example, $\leftindex_1{\bar{x}}_{k}^{my}$ can be the moving average for the past $3$ months, thus only involving values from the past ($\leftarrow$) of the same unit $k$ (unit).
    \item $\bar{\mathcal{F}}^{my}_{\leftarrow, \textrm{aggr}}$ contains features for each unit $k\in s^{my}$ constructed as an aggregated measure over a group which unit $k$ belongs to and using values only from the past ($\leftarrow$). For example, $\leftindex_{2}{\bar{x}}_{k}^{my}$ can be the $95th$ percentile of the moving averages $\leftindex_{1}{\bar{x}}_{k}^{my}$ over the economic activity class (4-digit classification code) which unit $k$ belongs to. 
    \item $\bar{\mathcal{F}}^{my}_{\updownarrow, \textrm{aggr}}(\tau)$ contains features for each unit $k\in s^{my}$ constructed as an aggregated measure over a group which unit $k$ belongs to and using values from the current time period ($\updownarrow$) at day $\tau$. For example, $\leftindex_{3}{\bar{x}}_{k}^{my}$ can be the $95th$ percentile of the turnover distribution of $z_{k}^{my}(\tau)$ over the economic activity class which unit $k$ belongs to. 
\end{itemize}

A detailed list of features and their coding for categorical variables is provided by \citet{BarBarCalGalMarRosSal22a}. Notice that at time $\tau$ features can be assigned for all units $k\in s^{my}$, since it does not require the values $z_{k}^{my}$. This set of features is completed with the target values $\{z_{k}^{my}(\tau_{f})\}_{k\in s^{my}}$ to constitute the data set $\mathcal{D}^{my}(\tau)$.\\

Next, we divide $\mathcal{D}^{my}(\tau)$ into the training data\footnote{We overload the notation by writing $(my)-k$ to denote the reference time period $k$ months before $(my)$. Notice that this is easily translated into many programming languages following the object-oriented paradigm.} $\mathcal{D}^{my}_{train}(\tau)=\bigcup_{\bar{m}\bar{y}\leq (my)-2}\mathcal{D}^{\bar{m}\bar{y}}(\tau)$ and test data $\mathcal{D}^{my}_{test}(\tau)=\mathcal{D}^{(my) -1}(\tau)$. We train a model on $\mathcal{D}^{my}_{train}(\tau)$, perform model selection on $\mathcal{D}^{my}_{test}(\tau)$, retrain the model on $\mathcal{D}^{my}_{train}\bigcup\mathcal{D}^{my}_{test}$ and apply it on $\mathcal{D}^{my}(\tau)$ to compute the predicted values $\{\hat{z}_{k}^{my}(\tau)\}_{k\in s^{my}}$. Details are given by \citet{BarBarCalGalMarRosSal22a}.\\

 The synthetic set $\bar{s}^{my}(\tau)=s^{my}(\tau)\bigcup\hat{s}^{my}(\tau)$ constitutes an anticipated version of the sample constructed using predicted values with the available information at time $\tau$ and data patterns from the past. This is the key idea: statistical offices potentially have enough microdata from short-term business statistics as to use statistical learning models to keep frequently updated anticipated versions of samples. These samples can now be used as in the traditional context to produce the corresponding aggregates.\\

For the Spanish Industrial Turnover Index, the sample is selected by cut-off \citep{INE_ICN18a} and the turnover total in population domain $U_{d}$ (geographic, sectorial, etc.) is estimated by $\widehat{Z}_{U_{d}}^{my}=\sum_{k\in s_{d}^{my}}z_{k}^{my}$, where $s_{d}^{my}\subset U_{d}^{my}$ is the cut-off sample. At any time $\tau$ this total can be early estimated by 

\begin{equation}\label{eq:TotalEstim}
\widehat{Z}_{U_{d}}^{my}(\tau)=\sum_{k\in \bar{s}_{d}^{my}}\bar{z}_{k}^{my}(\tau)=\sum_{k\in s_{d}^{my}(\tau)}z_{k}^{my}(\tau)+\sum_{k\in\hat{s}_{d}^{my}(\tau)}\hat{z}_{k}^{my}(\tau).
\end{equation}

At any time $\tau$ the total estimates \eqref{eq:TotalEstim} are ready to be used in the computation of econometric indices such as the fixed-base Laspeyres index used in the Spanish Industrial Turnover Index Survey \citep{INE_ICN18a}, producing thus early estimates of the same collection of indices and variation rates routinely produced and released monthly.\\

\begin{figure}[!htb]
\includegraphics[width=\textwidth]{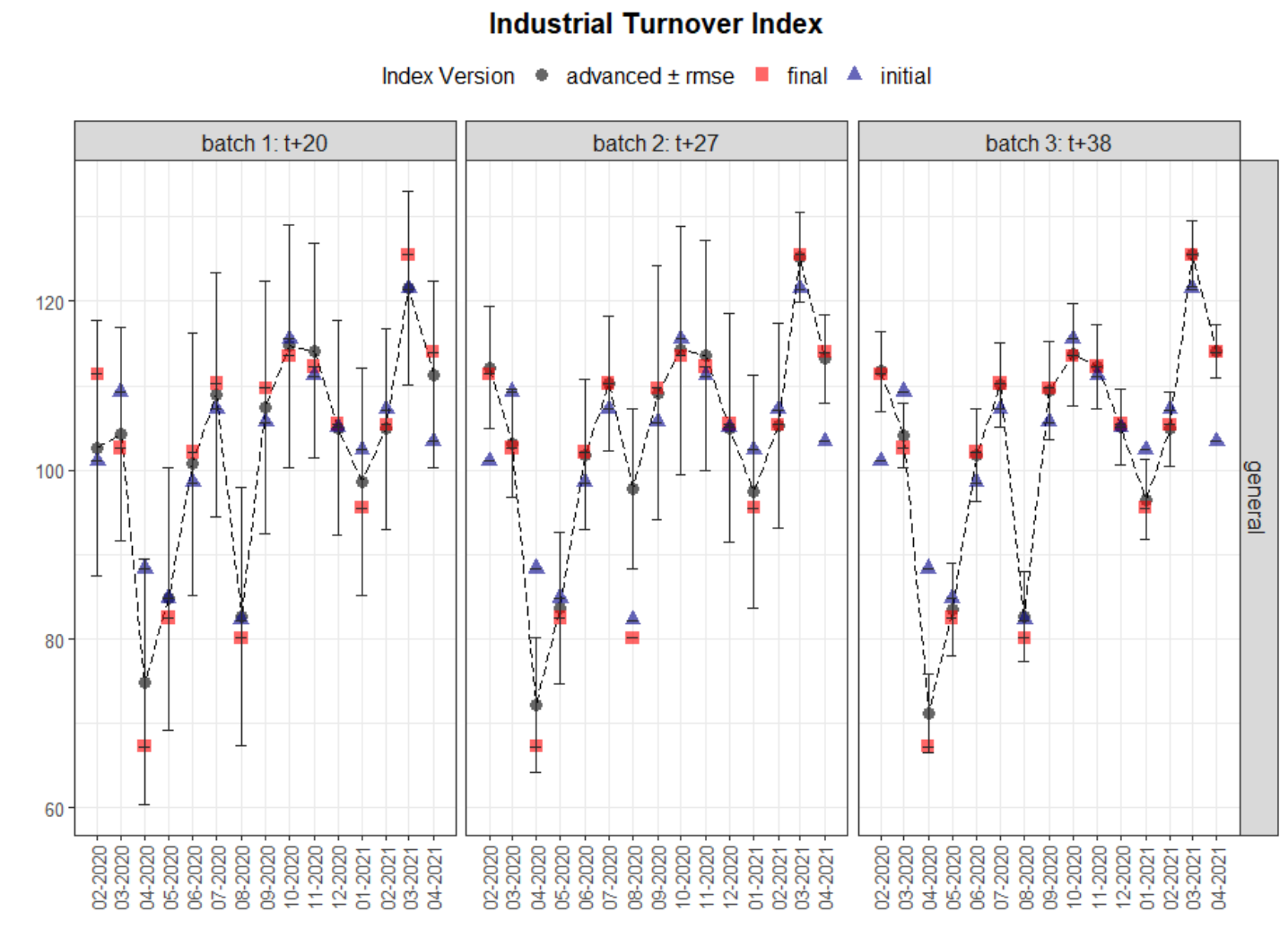}
\caption{\label{fig:Index} Early estimates of the national Spanish Industrial Turnover Index from Feb 2020 to Apr 2021.}
\end{figure}

\begin{figure}[!htb]
\includegraphics[width=\textwidth]{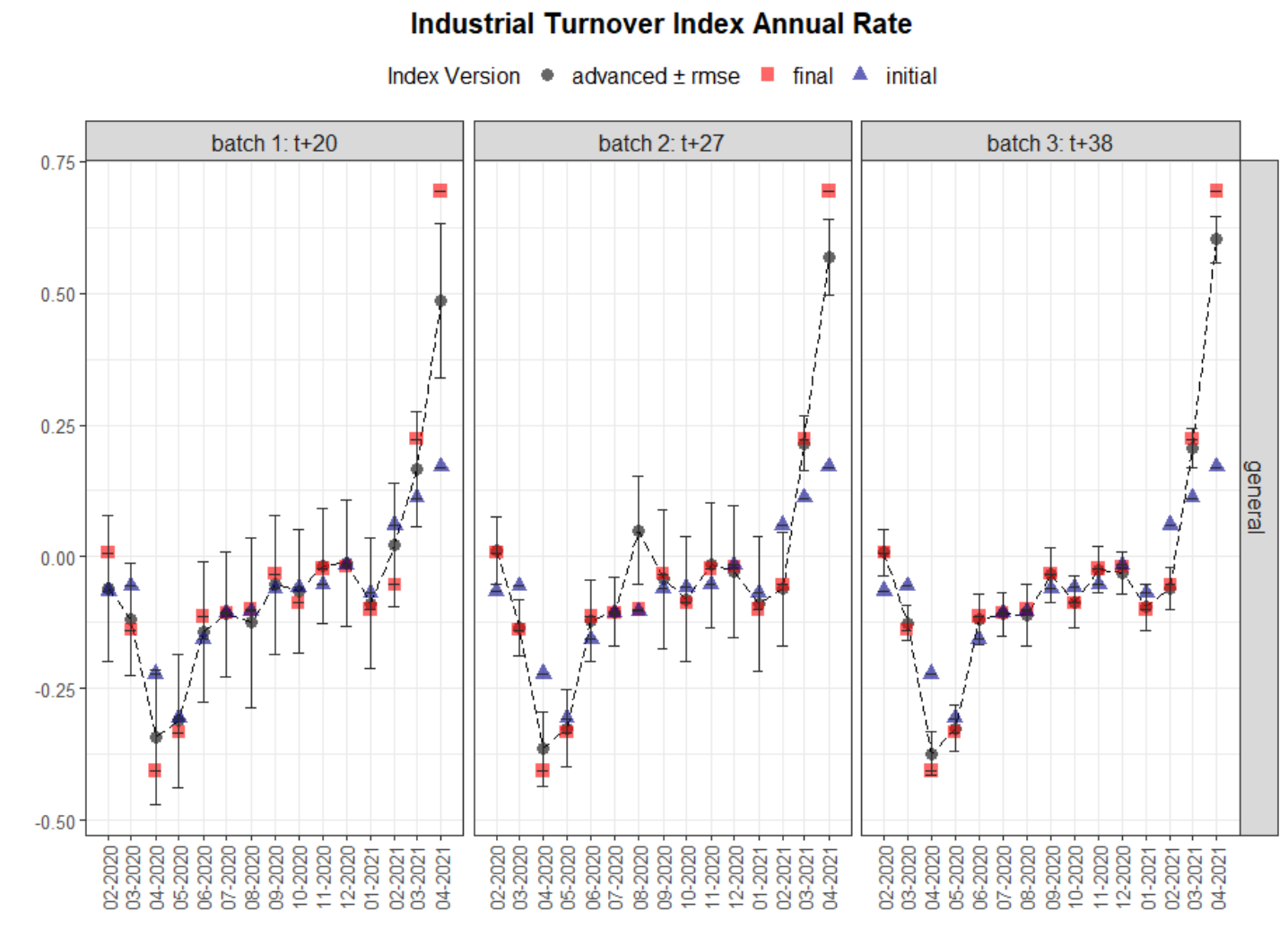}
\caption{\label{fig:Rate} Early estimates of the annual variation rates of the national Spanish Industrial Turnover Index from Feb 2020 to Apr 2021.}
\end{figure}

Our preliminary proof of concept focuses on a first implementation of the key idea above to construct the anticipated sample $s^{my}$ for every month $my$ over $5$ years and compare the early estimates with the final released values and with the predicted values not using cross-sectional data $\mathcal{F}_{\updownarrow}(\tau)$ from each current month (initial estimate) (see Figures \ref{fig:Index} and \ref{fig:Rate}).\\

Uncertainty in the predicted values for each unit and for the early aggregate estimates can be accounted for. Firstly, we focus on the conditional mean squared error, i.e.\ on the uncertainty for a given sample $s^{my}(\tau)$ regarding the model prediction for these units and this time period.  As a first immediate choice, neither do we assume independence and equal distribution nor exchangeability among the industrial establishments and estimate prediction errors at the current time period individually for each unit $k$ using prediction errors from the past \citep[see][]{BarBarCalGalMarRosSal22a}. The mean squared error for the total estimates \eqref{eq:TotalEstim} is computed by direct aggregation.  We then use this aggregate as a first immediate figure of merit to account for the uncertainty in the early estimate (see Figures \ref{fig:Index} and \ref{fig:Rate} for a graphical representation of this aggregate).\\

We comment on methodological and strategic aspects regarding this proposal and its implementation for official statistical production:

\begin{enumerate}
    \item The representation learning step is central in the prediction exercise. The construction of the features set $\mathcal{D}^{my}(\tau)$ from the original variables $\mathcal{F}^{my}(\tau)$ has been conducted in very close collaboration with domain experts to incorporate as much subject matter knowledge as possible (distribution tails, sectorial breakdowns, etc.) by defining and computing new features to incorporate into the prediction model. The possibility of using deep learning for this step remains to be explored.
    \item No special attention at all has been paid to model selection and hyperparameter optimization, which can and should be exhaustively improved in real production conditions. There is ample room for improvement in this course of action.
    \item Uncertainty has been elementarily accounted for, only minimally avoiding the identical distribution or exchangeability assumptions. More sophisticated uncertainty assessments must be explored (such as e.g.\ conformal prediction).
    \item In line with the avoidance of the identical distribution or exchangeability assumptions, outliers arise as an extraordinary challenge to predict, especially regarding large units, which strongly influence the indices. No special treatment has been explored and this needs to be addressed in real production conditions.
    \item The traditional production process needs some adjustments, for example, by making collected and validated microdata as early available for model updating as possible, moving from a batch-based perspective to a continuous updating of the systems.
    \item In this line, the microdata validation process is crucial for a high-quality algorithm training. Again, editing and validation business functions need to be accommodated to incorporate subject matter knowledge as automatically and early as possible. This will reduce the chance of model drift and model deterioration.
    \item If we can provide reliable predicted values for each statistical unit, we can reformulate the sample selection problem as the problem to select those units which allow us to maintain the model quality so that not every single unit must be required to provide a response every single time period. The goal of the sample selection is not the final aggregate estimate but the quality of the prediction model. Therefore, we could reduce response burden.
    \item Survey data do not play a central role in this exercise and a similar approach could be considered for administrative data. However, access to these data needs to be restated. Usually, statistical offices have access to these data after the reference period is over and after some pre-processing and revision tasks are concluded by their holders. Data access and use is usually approached in batches. Timeliness could be improved should offices have earlier and continuous access to administrative data as soon as they are generated and/or collected.
    
\end{enumerate}

\subsection{Imputation beyond the sample}
\label{sub:ImputBeySample}

One of the current demands from the users of official statistics is the need to improve the granularity of the statistics, providing more disaggregated information while maintaining a good quality. Moreover, there is also the need to reduce the response burden and costs of the operation. Clearly, there arises a trade-off. Most efforts to incorporate new data sources into the production of official statistics aim at seeking a solution for the trade-off. In order to achieve these goals, we need to leverage the large amount of administrative data available nowadays in order to provide better estimates with a smaller sample.\\

Our proposal consists in constructing the microdata for all the population. Acknowledging the highly predictive power of statistical learning models, especially in data-rich environments such as those arising from the access to administrative registers, we shall use Machine Learning techniques to impute the values of the variables of interest for the entire target population. While this construction of population microdata has several potential uses, in this section we will focus on its use for the estimation of population totals. By using Machine Learning models, we hope to make use of the auxiliary information in a most accurate way to obtain good quality microdata for all the units in the population, and hence to be able to provide high quality estimates at a finer degree of granularity.\\

Let us describe the basic idea behind our proposal. Suppose we have some population $U$ and for each unit $k \in U$ we have available some auxiliary data $\mathbf{x}_k$ (for instance, administrative data) and some variables of interest $y_k$ which we want to impute for all the population\footnote{As explained below, imputed values for units in the sample with known value of $y_{k}$ are used for better estimation performance.}. Suppose moreover that we have available a sample $s$ obtained from $U$ by probability sampling, and that we know the value of $y_k$ for $k \in s$. Then, we can proceed as follows. We use the sample $s$ to select and train some model $\mu(\mathbf{x},s)$ for the target variable $y$ in terms of the auxiliary data $\mathbf{x}$. Then, for each unit $k$ in the whole population we can impute the value of $y_k$ by $\hat{y}_k = \mu(\mathbf{x}_k,s)$. Finally, suppose we want to provide an estimate of the population total of the variable $y$. One can do it straightforwardly by using a prediction estimator, which just uses the imputed values outside the sample:
$$\hat{Y}^{pred} = \sum_{k \in s} y_k + \sum_{k \in U \setminus s} \mu(\mathbf{x}_k,s).$$

There are at least two issues that arise in this naive approach. First, there is the problem of how to choose an appropriate model for each target variable. This entails both deciding the kind of model to be used (linear models, tree-based models like random forest or XGBoost, neural networks,  ...) together with its hyperparameters, and training the model. It is important to note that, in order to give the best possible estimate for the population total, we should select the model that minimizes the squared total error
$$E^2 = (Y - \hat{Y})^2,$$
which will not necessarily agree with the model which gives the more accurate individual predictions. Another important issue that needs to be addressed in this approach is that of uncertainty quantification. We want not just an estimate for the population total, but also a measure of the quality of that estimate (for instance, an estimate of the $MSE$ of the total estimator).\\

Both problems are solved by using the algorithm-assisted\footnote{We use the term \emph{algorithm-assisted} in the sense of generalising in a natural way the ideas of \emph{model-assisted} estimation \citep[see e.g.][]{SSW92}.} estimation paradigm described in Section~\ref{sub:AlgAssEst}. As described there, we can make use of the sample $s$ in order to select and train the appropriate model. The same paradigm also provides us with an unbiased estimate of the MSE of the total estimator.\\

In some cases, we could even use prediction estimators based on pretrained models (for instance, models selected and trained in a sample of the previous year) to provide predictions of the population totals for the variables of interest. In this case, if $\mu(\mathbf{x})$ is our pretrained model for $y$, the prediction estimator would be given by
$$\hat{Y}^{pred} = \sum_{k\in U} \mu(\mathbf{x}_k).$$

This would allow us to obtain estimates for the population totals without using any sample. However, one has to keep in mind that this approach has several drawbacks:
\begin{enumerate}
    \item We expect some model drift as time passes, meaning that the model performance degrades over time. By using the same model at different points in time we are assuming the hypothesis that the model is not changing in a significant way. Moreover, if we have no available sample at the current time we have no way to assess if this hypothesis is true.
    \item If we have no sample, we have no way to obtain a reliable quality indicator of the estimates. Again, we must rely on the assumption that the quality is similar to that of the last time where we had some sample.
    \item In the situation where we have an available sample, we can use the values of the sampled units in our estimation. This can improve substantially our estimation, especially in the presence of an exhaustive stratum which contains the most influential units. On the other hand, when we have no sample we have to rely on the imputed values for all the units of the population.
\end{enumerate}

It is also important to remark that, while we are mostly ignoring this issue here, the quality of the auxiliary data is very important in order to guarantee good estimates. Therefore, some quality indicators of the auxiliary data should also be given (e.g.\ indicators about their completeness, coverage, validity, timeliness,  ...).\\

As a proof of concept, we explain how we apply these ideas to the Spanish Structural Business Statistics (SBS). The goal of the SBS is to provide information on the main structural and economic characteristics of the enterprises, in the different sectors studied, through a wide range of variables relating to the personnel employed, turnover and other incomes, purchases and consumption, personnel expenditure, tax and investment. This information is given separately for each of the following sectors: industry, trade and services. The information is provided both for the totality of the sector and also disaggregated by NACE and by region. Currently, the sampling design used is stratified sampling, and the estimates of the variables are computed via Horvitz-Thompson (HT) estimators with calibration to the population totals of some auxiliary data. See \citet{INE_SBS21} for details.\\

Our objective is to use imputation beyond the sample using ML methods in order to significantly reduce the sample size or, if possible, even eliminate the sample in some years, while maintaining or improving the current quality of the statistics. More precisely, we want to estimate the population totals of $95$ quantitative survey variables, using as auxiliary data $150$ variables coming from administrative data of the Spanish and Statutory Tax Agencies. While the goal is to provide estimates at several levels of disaggregation, in this description we will focus just on the aggregated totals for all the economy.\\

As the first step of the project, we have dealt with a worst-case scenario. We assume that we have no sample for the current year and that the only available sample is that of the previous year. Moreover, the year chosen for our study is $2020$ (the COVID19 pandemic year) using sample from the year $2019$. Therefore, it is reasonable to expect that the model drifting effect will be more significant than in any other pair of consecutive years. As models for our target variables, we try a variety of Machine Learning models (linear regression with elastic net regularization, random forests and XGBoost) with hyperparameter tuning, and we make both model selection and training using exclusively the sample from the year $2019$, therefore using the models in $2020$ as pretrained prediction estimators. It is also important to remark that in training the models with the $2019$ sample we use the sampling weights, so that the different importance of each sampled unit can be taken into account in the model.\\ 

\begin{figure}[!htb]
\centering
\includegraphics[width=0.5\textwidth]{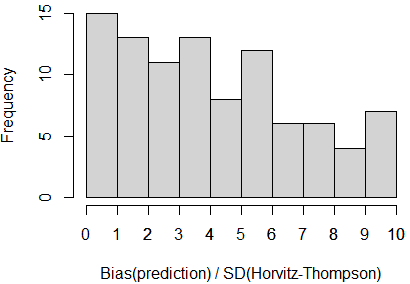}
\caption{\label{fig:SBShist} Histogram of the relative efficiency of the prediction estimators with respect to the Horvitz-Thompson estimators for the $95$ target variables.}
\end{figure}

We have obtained estimates of the $95$ variables of interest, both using the pretrained prediction estimators and the currently used HT estimators with the sample of $2020$, and we have compared the accuracy of the two estimators using the sample for $2020$. Let $\hat{Y}^{pred}$ be the pretrained prediction estimator of one variable, and $\hat{Y}^{HT}$ the HT estimator of the same variable. In order to compare the accuracy of both estimators we use the usual estimator of the variance for the HT estimator, $\hat{V}(\hat{Y}^{HT})$, and we estimate the bias of $\hat{Y}^{pred}$ by using a Horvitz-Thompson-like estimator upon the $2020$ sample:
$$\hat{B}(\hat{Y}^{pred}) = \sum_{k\in s_{2020}} \frac{\hat{y}_k - y_k}{\pi_k},$$
where $s_{2020}$ stands for the sample in the year $2020$, $\hat{y}_k$ is the prediction of the pretrained model for the unit $k$, and $\pi_k$ are the sampling weights for $2020$. Observe that, since the pretrained estimators do not depend on the chosen sample for the year 2020, they have zero sampling variance but are generally biased estimators. On the other hand, HT estimators are unbiased but have nonzero sampling variance \citep[see e.g.][]{SSW92}. Therefore,
$$\frac{\widehat{rmse}(\hat{Y}^{pred})}{\widehat{rmse}(\hat{Y}^{HT})} = \frac{\hat{B}({Y}^{pred})}{\sqrt{\hat{V}(\hat{Y}^{HT})}}$$
provides an estimate of the relative efficiency of $\hat{Y}^{pred}$ with respect to $\hat{Y}^{HT}$. The prediction estimator is more efficient than the HT estimator when this quotient is less than $1$. The histogram of this relative efficiency for the $95$ target variables can be seen in Figure~\ref{fig:SBShist}. \\

We see that only for $15$ of the $95$ target variables the accuracy of the prediction estimator is better than that of the Horvitz-Thompson estimator. This means that, at least for $2020$, the models pretrained with $2019$ data in general don't provide enough accuracy to replace the HT estimators. However, it is noteworthy that even in this worst-case scenario there are some variables for which the prediction estimator outperforms the HT estimator. From the results obtained thus far, we can conclude that some sample will be necessary each year in order to deal with model drift. Moreover, a well-chosen sample will very likely improve substantially the quality of the estimates.\\

We make some additional comments and remarks.

\begin{enumerate}
    \item For some units there is essentially no auxiliary data available, so for those units the imputed values using models are unreliable. On the other hand, there are some big units which have a significant contribution to the total, and knowing the value of those units (by surveying them each year) would improve the quality of the total estimation. It is important to notice that even statistical learning methods fall short of providing a working solution at this point.
    \item There are some target variables which are essentially independent from the available auxiliary data. Therefore, for these variables it will be impossible to obtain good predictions using models based only on the auxiliary data. Again we notice that even statistical learning methods fall short of providing a working solution at this point.
    \item No regressor selection has been made thus far. With well-chosen regressors for each model we expect to improve the quality of the predictions, and hence the quality of the total estimates. This situation is similar to that in early imputation for constructing the features of the model. The representation learning step in the use of these models constitutes the next step. In this regard, the collaboration with the business experts in order to define the best regressors for each target variable is indispensable.
    \item The model selection process has not been very exhaustive, mainly due to the fact that this is the most computationally expensive part of the process. A more exhaustive model selection process should improve the estimates. As in early estimation (and use cases in the next sections), we have made an educated choice of the models, which now should be successively improved, potentially with an exhaustive search, but perhaps more sustainably with a continuously evolving process.
    \item As mentioned before, in this exercise we have used the sample of $2019$ for the model selection and to train the models, while we use the sample of $2020$ just to compute an estimate of the accuracy of our total estimates. In a situation where we assume we have a sample in the same year we want to make the estimation, we would use the same sample for model selection, training, and estimation of the MSE. We would also use the known values for the units in the sample instead of the predictions in the estimation of the total.
\end{enumerate}

To conclude, we note that the next step is the determination of a reduced sample which allows us to significantly reduce the sample size while maintaining the quality using algorithm-assisted estimation. Given the characteristics of the Structural Business Survey, it seems that the better approach would be to survey the bigger enterprises, which contribute substantially to the population totals, survey also the units with no auxiliary information, and finally select a well-chosen sample within the companies of intermediate size. In designing this sample the expertise of business experts and sampling experts will also be essential. The sample thus selected will be aiming at excelling model performance even for highly granular information. In our view, even when Machine Learning methods are used, surveys are not recommended to be discarded in order to achieve accurate estimates.

\subsection{Integration of administrative data as a primary source in business statistics}
\label{sub:IntDatSour}

The proposal for the production framework using administrative data as the primary source begins from the hypothesis of maintaining the same objectives as in the case of survey data, i.e., the aim is to estimate a set of population aggregates in a finite population \(U\), defined as \(Y_{U_{d}}=\sum_{k\in U_{d}}f(\mathbf{y}; \mathbf{x})\) for a collection of population domains \(U_{d}\subset U\) (publication cells) for various target variables \(\mathbf{y}\) and auxiliary variables  \(\mathbf{x}\). Without loss of practical generality, we can focus on population totals in the form \(Y_{U_{d}}=\sum_{k\in U_{d}}y_{k}\), as other more complex aggregates can be expressed as functions of these totals. We have the complete sample \(s=\bigcup_d s_d\), where \(s_{d}\subset U_{d}\).\\

To achieve this, similarly to survey data, we will analyze the use of linear estimators of the form \(\widehat{Y}_{U_{d}}=\sum_{k\in s_{d}}\omega_{ks}(\mathbf{x})y_{k}^{\circ}\), \(\omega_{ks}(\mathbf{x})\) are pseudo-sampling weights (or true weights if a sampling design is used), and \(y^{\circ}\) denotes a synthetic value for variable \(y\), which can either be a transformation of the corresponding administrative variable or a predicted value for the survey variable based on all available information (administrative and survey). Accuracy measures must also be produced.\\

In this context, the fundamental concepts of finite population and target variable remain to be central. In consequence, for quality assurance the paradigm of the total survey error model \citep{GroLyb10a} is still valid, even under its consideration as the second phase of the two-phase life-cycle model by \citet{Zha12a}. We shall focus on short-term business statistics incorporating tax register data as primary data source in combination with survey data under a given probabilistic sampling design. In particular, we shall provide the description of an ongoing pilot experience with the Service Sector Activity Indicators (SSAI) survey, which is beginning to use VAT data from the National Tax Agency to reduce response burden.\\

Let us denote by $U^{adm}$ the set of business units contained in the tax register and by $U$ the finite population of analysis, which is represented by the population frame $U_{F}$ obtained from the complete business register in our office. Our first concern is about the coverage error, particularly on identifying those administrative units $k\in U^{adm}$ considered as statistical units $k\in U$. From the tax register $U^{adm}$ we shall consider statistical units $k\in U$ those units $k$ also contained in the frame population, i.e. $k\in U^{adm}\cap U_{F}$. The target statistical variable for these units will be synthetized using the raw administrative value $y^{adm}$ in a model. We shall denote $U^{mdl}=U^{adm}\cap U_{F}$.\\

A first natural course of action is to use the administrative values \(y^{adm}\) by mere substitution as the value of the statistical values \(y^{stat}\). This would be a more cost-efficient and timely strategy \emph{provided the quality of the input administrative data is guaranteed}. In this sense, taking advantage of having both the administrative and statistical values for a subset of units $k\in s\cap U^{mld}$, we assess the quality of the input tax data for statistical purposes as follows.\\

Several proposals in official statistics aim to measure quality at various stages of the statistical production process. However, historically, more emphasis has been placed on assessing the quality of final aggregates rather than on the input data, primarily due to the control inherent in the generation of survey data. With the growing incorporation of diverse data sources, there is an increasing need to evaluate their quality as well. Initiatives within the European Statistical System (ESS), such as the BLUE-ETS Project \citep{daas2011} and the ESSnet KOMUSO \citep[see][and references therein]{KOMUSO} have addressed this need.\\

\begin{figure}[!htb]
\begin{center}
\includegraphics[width=0.7\textwidth]{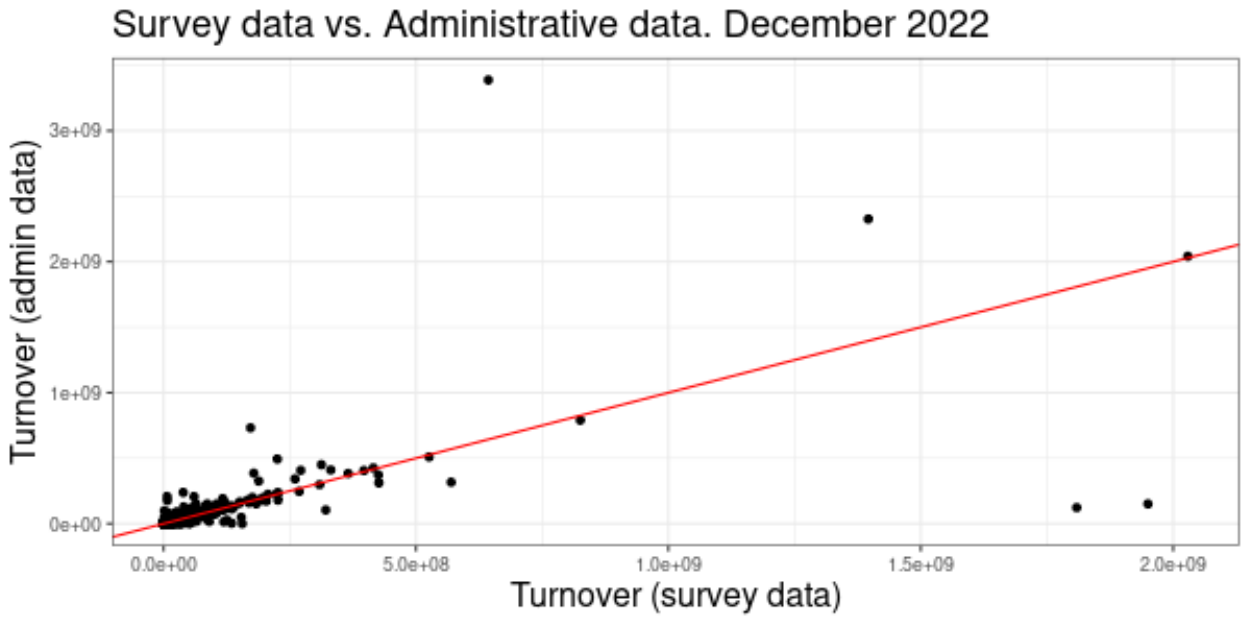}
\caption{\label{fig:micro_admin-surv} Microdata comparison between survey and administrative data.}
\end{center}
\end{figure}

At Statistics Spain, motivated by the 3rd round of Peer Reviews of the European Statistical System \citep{ESSPR3}, we have recently started to engage in the development of a diverse set of indicators aimed at evaluating the quality of data sources across multiple dimensions \citep{NieBarRodSalSal24a}. Furthermore, we are conducting retrospective analyses to juxtapose administrative data with survey data at the microdata level. This endeavor entails the creation of numerical indicators to facilitate a comprehensive and rigorous evaluation process for their usage as input data in official statistical production. Meanwhile, a graphical representation of this comparison can be seen in Figure~\ref{fig:micro_admin-surv} where the turnover variable is represented for the units in common between administrative and survey data for reference time period of year 2022.\\

\begin{figure}[!htb]
\begin{center}
\includegraphics[width=0.7\textwidth]{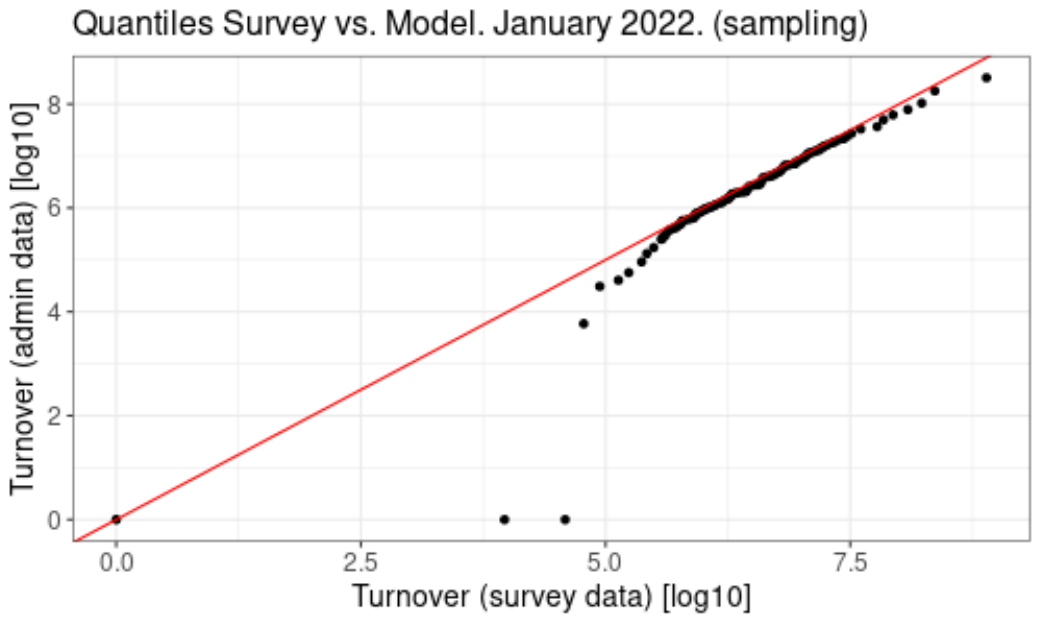}
\caption{\label{fig:QQmues_admin} Quantile comparison between survey and administrative data for January 2022 (sample level).}
\end{center}
\end{figure}

This rule-of-thumb comparison has been complemented focusing on the distribution of both datasets (admin and survey) for the target variable. This is relevant to understand where the differences lie.
Graphical representations depicting the quantiles corresponding to each dataset have been generated. Figure~\ref{fig:QQmues_admin} represents the sample quantiles of the survey data against those of the administrative data for the January 2022 period. This makes clearly explicit the issues with both distribution tails.\\ 

\begin{figure}[!htb]
\begin{center}
\includegraphics[width=0.7\textwidth]{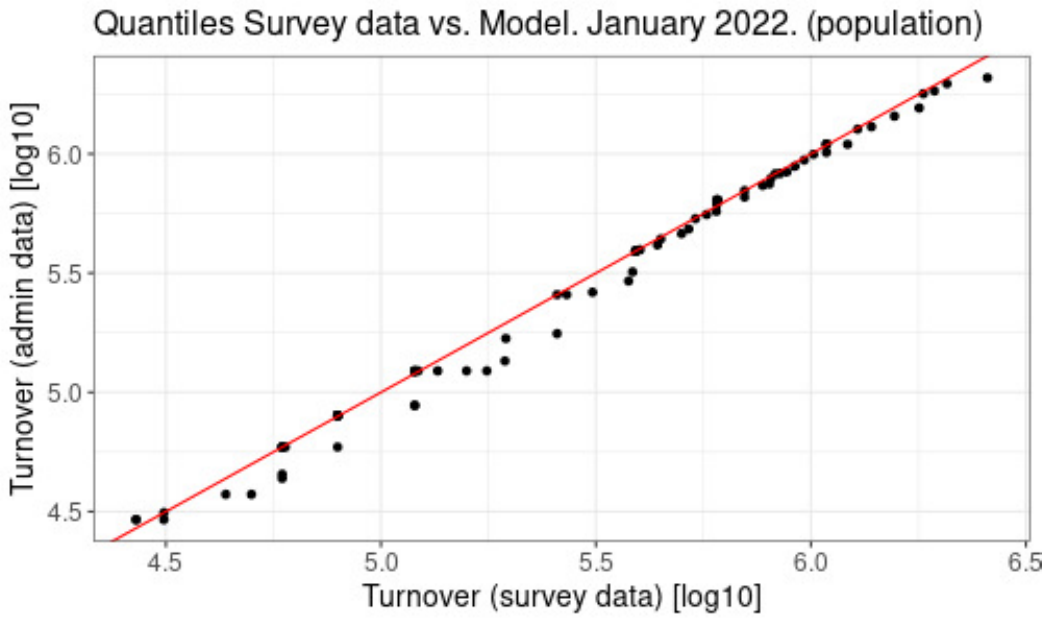}
\caption{\label{fig:QQpob_admin} Quantile comparison between survey and administrative data for January 2022 (population level).}
\end{center}
\end{figure}

The comparison can be conducted also upon the population-level estimates. In particular, in Figure~\ref{fig:QQpob_admin} we represent the estimated population quantiles for the same period, thus making visible the effect of the sampling weights. As these representations show, the differences between both sources are significant at microdata level. The discrepancy is not as pronounced concerning the estimated population distributions as evidenced by the comparison in Figure~\ref{fig:QQpob_admin}. However, these differences become greater as we focus on more disaggregated population domains $U_{d}$, thus impinging negatively on granularity.\\

As a consequence, our next concerns are the validity and measurement errors. It is well known \citep{KOMUSO} that administrative data can severely differ from survey data since the latter are defined and collected for statistical purposes. In this sense, in the particular case of the SSAI survey, the administrative total sales value  \(y^{adm}\) declared for tax purposes may differ from the statistical total turnover value \(y^{stat}\) traditionally collected in questionnaires. These differences cautiously discourage the use of \(y^{adm}\) by mere substitution as the value of \(y^{stat}\).\\ 

Our proposal aims at a combined use of statistical learning models and data validation techniques to control this difference (validity error, but also measurement error since we use validated microdata as training data). Longitudinal information is of special relevance as auxiliary information. Consider several datasets for reference time periods \(t_{-1}\), \(t_{-2}\), \(t_{-3}\), and so on, where past periods \(t_{-i}\) will be used for model training. The proposal focuses on predicting and validating successively each dataset \(t_{0}\), \(t_{1}\), \(t_{2}\), etc. with their past datasets.\\

Firstly, to initialise the recurrent modelling exercise in successive time periods, training sets for $t_{0}$ are identified as those units in the probabilistic samples and the tax register, i.e. $k\in s_{t_{-i}}^{mdl}=s_{t_{-i}}\cap U^{mdl}$. Their corresponding synthetic target variable values \(y^{\circ}_{k}\) are the validated values entering the computation of the indices, i.e.\ \(y_{kt_{-i}}^{\circ}=y_{kt_{-i}}^{stat}\). Then, a statistical learning model \(y^{\circ}=f(y^{adm};\mathbf{x}) + \epsilon\) is adjusted using explicitly the value of the administrative variable as a feature. Once the model is constructed, it is used to predict the values of the variable \(\widehat{y}_{kt_{0}}=\widehat{f}(y_{k}^{adm}; \mathbf{x}_{k})\). Notice that this is the predicted value of the validated total turnover in terms of the raw administrative value of the total sales variable (and other features). The rest of features $\mathbf{x}$ are constructed as in the early imputation section for the early estimates of the ITI (cf.\ Section~\ref{sub:EarlyImput}). These predicted values are candidates to enter the index computation.\\

Next, a data validation strategy is applied. This involves designing and applying both error detection functions (edits) and their treatment, which likely requires a more specific imputation model. Ideally, this part should be automated to the highest possible degree. The result will be a new refined set of validated synthetic target values together with the validated survey data values $s_{t_{0}}$ used for index computation.\\

The main objective of using administrative data as the primary source is to reduce the burden on respondents. To achieve this, questionnaires should be eliminated for those units with reliable administrative information. However, the selection of such units must be carried out carefully, as there may be a lack of information to properly train models that predict values for units not included in the survey sample.\\

Many scenarios based upon scoring to select units reporting survey data and units reporting with their administrative records have been tested. The one presented here has been yielding the best results so far. In this scenario, specific criteria have been defined to identify units exhibiting erratic behaviour for survey reporting, thereby ensuring the quality of their values through the traditional data collection and data editing processes. Survey-reporting units satisfy at least one of the following criteria.\\

\begin{figure}[!htb]
\begin{center}
\includegraphics[width=0.75\textwidth]{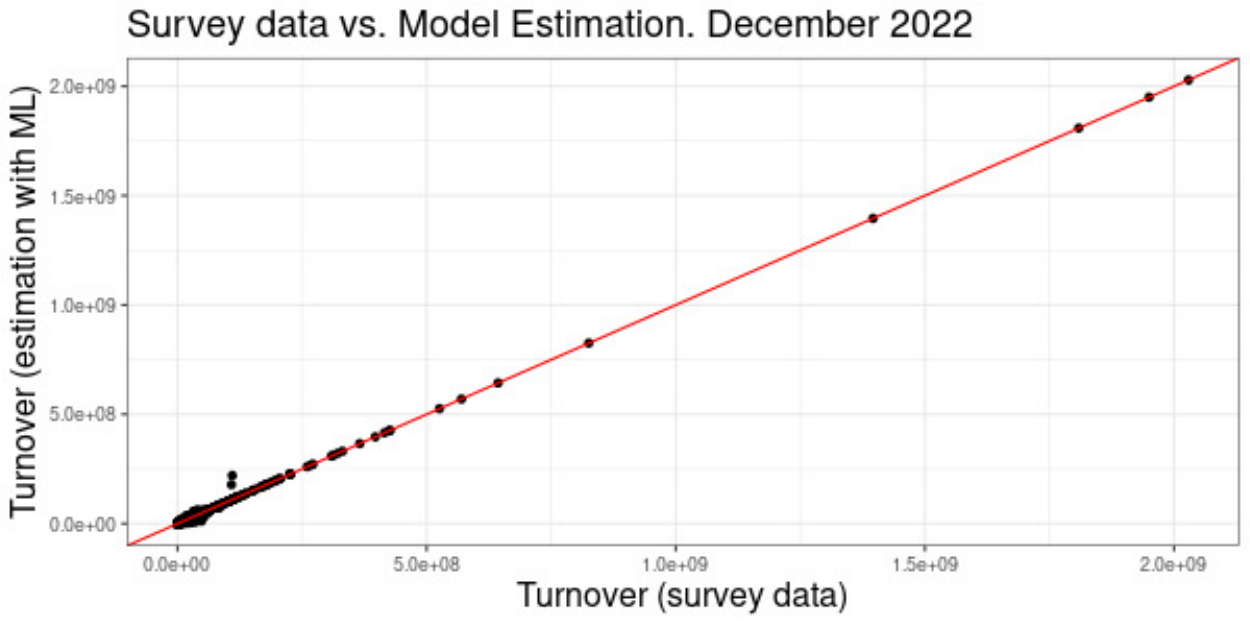}
\caption{\label{fig:micro_model-survey} Comparison between survey microdata and model-predicted microdata.}
\end{center}
\end{figure}

\begin{description}
    \item[\underline{Criterion 1}] Units with a high impact on the aggregate.\\ 
    A first period-wise score for unit $k$ in period $t_{-i}$ is defined as $r_{kt_{-i}} = \frac{\omega_{kt_{-i}} y_{kt_{-i}}^{stat}}{\widehat{Y}_{dt_{-i}}}$, where $\widehat{Y}_{dt_{-i}}$ is the aggregate estimator for domain $d$ in period $t_{-i}$. The periods corresponding to the first nine months of the previous year to the reference year are used, and a first global score $r_{k}^{(1)}$ for each unit $k$ is defined as the $p$th quantile $r_{k}^{(1)} = Q_{p}(r_{kt_{-1}}, \dots, r_{kt_{-9}})$. We have used the median ($p=0.5$). A threshold is computed using a conservative elbow criterion \citep{SatAlbIrwRag11a}, and, thus, units above the threshold are selected for survey-reporting.
    \item[\underline{Criterion 2}] New units.\\
    All new units in the sample $s$ of the previous year to the reference time period that have $\omega_{kt_{-i}}=1$ or annual turnover in frame $U_{F}$ for the preceding year greater than a chosen threshold $t_{F}$ are selected. We have used $t_F = 10^{7}$. A second global score $r_{k}^{(2)}$ is thus defined as $r_{k}^{(2)} = I\{\omega_{kt_{-i}}=1 \lor a_F> 1e7\}$. Notice that $\omega_{kt_{-i}} = \omega_{kt_{-j}}$ for all $t_{-i}$ and $t_{-j}$ in the same year since sampling designs change only annually. Units with $r_{k}^{(2)}=1$ are selected for survey-reporting.
    \item[\underline{Criterion 3}] Units with high variability in the target variable.\\
    Another global score is defined as $r_{k}^{(3)} = \sqrt{Var(\omega_{kt_{-1}}y_{kt_{-1}}, \dots, \omega_{kt_{-9}}y_{kt_{-9}})}$. Again an elbow-based threshold is used to select those units for survey-reporting.\\  
    \item[\underline{Criterion 4}] High difference between the survey and administrative values.\\
    A time-wise score for unit $k$ in period $t_{-i}$ is defined as $r_{kt_{-i}}^{(4)} = \frac{\omega_{kt_{-i}} |y_{kt_{-i}}^{adm} - y_{kt_{-i}}^{stat}|}{\widehat{Y}_{dt_{-i}}}$.  A new global score is defined as $r_{k}^{(4)} = \sqrt{Var(r_{kt_{-1}}^{(4)}, \dots, r_{kt_{-9}}^{(4)})}$. Again an elbow-based threshold is used to select those units for survey-reporting.\\
    \item[\underline{Criterion 5}] High absolute differences between the survey value and the administrative record value.\\
    Using the same time-wise score for unit $k$ in period $t_{-i}$ as in Criterion 4, a new global score is defined as the $p$th quantile so that $r_{k}^{(5)} = Q_{p}(r_{kt_{-1}}^{(4)}, \dots, r_{kt_{-9}}^{(4)})$. We have selected $p=0.5$. Again an elbow-based threshold is used to select those units for survey-reporting.\\
    \item[\underline{Criterion 6}] Zero values.\\
    All units with any administrative record value equal to zero in the periods under consideration are selected. The global score is defined as $r_{k}^{(6)} = I\{y_{kt_{-1}}^{adm}=0 \lor y_{kt_{-2}}^{adm}=0 \lor \dots \lor y_{kt_{-9}}^{adm}=0\}$. Units with $r_{k}^{(6)}= 1$ are selected for survey-reporting.
 \end{description}

Units not selected according these scoring system will be modelled. These criteria are conservative with respect to reduction of response burden with the idea of keeping all challenging units in the survey. For example, for year 2021, there were 5,281 units in the administrative dataset, and using these criteria, still 2,320 would be needed to collect in the survey and 2,961 could be dropped.\\

As preliminary results, again we can report both at the statistical unit level and at the aggregate level. Firstly, in Figure~\ref{fig:micro_model-survey} we represent the predicted values for the total turnover values in comparison with the survey value, which is equivalent to Figure~\ref{fig:micro_admin-surv}. In this case the comparison is obtained after implementing the proposed model. A substantial improvement is now evident in how closely the model values resemble those of the survey, thus partially reproducing the data editing tasks to account for validity and/or meausurement errors.\\

\begin{figure}[!htb]
\begin{center}
\includegraphics[width=0.80\textwidth]{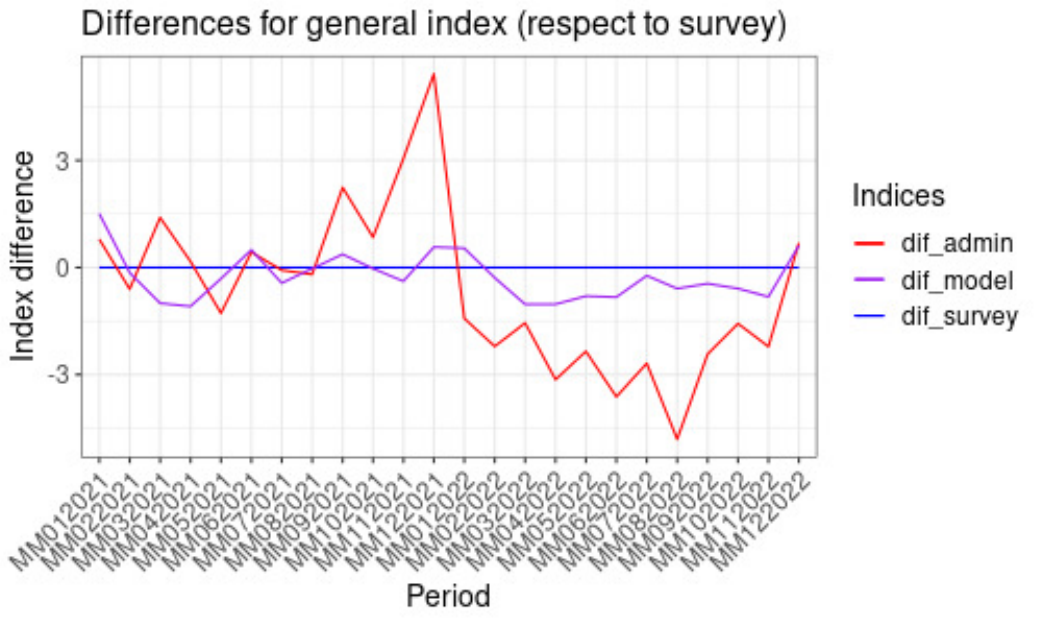}
\caption{\label{fig:dif_gral} Differences between the general index of SSAI obtained with administrative and the model respecting to only survey data.}
\end{center}
\end{figure}

At the aggregate level, Figure~\ref{fig:dif_gral} presents a comparison of the SSAI index obtained through direct substitution of administrative data for those units intersecting with the sample (dotted line) and the index obtained using model-predicted and validated values (dashed line). The graph illustrates the differences between each of these indices and the index obtained exclusively from survey data (solid line). It can be observed how the differences are considerably smoothed out by using a model to account for validity and measurement errors in the statistical value based on the administrative value.\\

As main remarks we state that:

\begin{enumerate}
    \item To account for validity and measurement errors in administrative data using statistical learning models aiming at response burden reduction, a selection of units maintaining the model quality is advised.
    \item The selection problem is thus focused on the quality of the model, not on the quality of the final output, which is obtained by standard aggregation procedures.
    \item More research is needed to find a trade-off between accuracy (lack of measurement errors) and response burden reduction. Our guess is that the higher the predictive power of the model, the higher the reduction.
    \item Ongoing work is under development to estimate variances and mean squared errors arising from the combination of sampling designs and model predictions.
\end{enumerate}

\subsection{Time disaggregation of sampling designs}
\label{sub:TimDisSampDes}

Many official statistics collect data continuously (e.g. weekly) and process and aggregate them to produce and release monthly, quarterly, or even annual outputs. The whole production process follows the traditional survey methodology, with the selection of a probabilistic sample from a population frame, the use of well-known data collection modes integrating the interview administration, the data entry and the data editing during collection (e.g. using a CAPI data collection mode), the execution of a statistical data editing and imputation strategy, the calibration of sampling weights, the computation of the estimates and of coefficients of variation, and the preparation of the  dissemination products (including statistical disclosure control), to name the most significant.\\

This process is usually deemed expensive and slow, negatively impinging on timeliness \citep[see e.g.][]{Eur17a}. Here we show how the introduction of novel methods in this process allows us to take advantage of the continuous data collection to produce more frequent estimates. We shall use the Spanish Labour Force Survey to illustrate our proposal. The starting point is the usual linear estimator formula for constructing the statistical outputs: $\hat{A}_{d}(C)=\sum_{k\in s_{d}}\omega_{ks}(\mathbf{x})\delta_{k}(C)$, where $A_{d}(C)=\sum_{k\in U_{d}}\delta_{k}(C)$ denotes the $C$-property class total in the population domain $U_{d}$, $\omega_{ks}(\mathbf{x})$ denotes the (usually calibrated with respect to marginal $\mathbf{X}$) sampling weight for unit $k$ and $\delta_{k}(C)$ is the binary indicator variable of property $C$ for unit $k$ (e.g. employed or unemployed).\\

We make the following considerations regarding time. Firstly, we observe two time scales in the survey. On the one hand, aggregates $A_{d}(C)$ and their estimators $\widehat{A}_{d}(C)$ are referred to a long time period such as quarters in a year (say, $\hat{A}_{d}^{Q}(C) $, with $Q=1, 2, 3, 4$). On the other hand, collected target variables $\delta_{k}(C)$ are referred to specific weeks\footnote{Usually the interview is administered one week after the reference week.} in a year (say, $\delta^{W}_{k}(C)$, with $W=1,\dots, 52$). These two time scales are connected through the sampling weights derived from the sampling design over the quarterly target population\footnote{In household surveys this is usually the population at the middle date of the time span period coming from the Population Register.}. In this connection the underlying assumption that the weekly collected value $\delta_{k}(C)$ is valid throughout the whole reference period $Q$ is made. This is a kind of ergodic hypothesis assuming that a cross-sectional behaviour in a given week can be assigned longitudinally to the whole quarter. This hypothesis is made to make the sampling size valid for the whole reference time period (e.g. the quarter).\\

Instead, we avoid this ergodic hypothesis and use a random forest algorithm to disaggregate the sampling designs from the quarterly scale down to the weekly scale. The reasoning is as follows. Let us denote by $\pi_{k}^{Q}$ the first-order inclusion probability for unit $k\in U$ according to the quarterly sampling design in production. As a production step in the whole process, every sampled unit $k\in s$ follows an assignment procedure in the quarter to administer the interview in the corresponding reference week $W$. This is accomplished semi-manually by experts accounting for a strict balance across the whole national territory for data collection fieldwork reasons. This week assignment is conducted taking into account frame and design variables such as regions, provinces, and strata. The quarterly sample $s^{Q}$ can then be disjointly partitioned into $13$ weekly samples $s^{Q}=\bigcup_{W=1}^{13}s^{W}$, $s^{W}\cap s^{W'}=\emptyset$, $W\neq W'\in\{1,\dots, 13\}$. We define the weekly first-order inclusion probabilities as $\pi_{k}^{W} = \mathbb{P}\left(s^{W}\ni k\right)$ and compute them using compound probability properties:

\begin{align*}
        \pi_{k}^{W} & =\mathbb{P}\left(s^{W}\ni k\right)\\
                    & =\mathbb{P}\left((s^{Q}\ni k) \land (k\rightsquigarrow W)\right)\\
                    & =\mathbb{P}\left((k\rightsquigarrow W)|(s^{Q}\ni k)\right)\cdot\mathbb{P}\left(s^{Q}\ni k\right)\\
                    & = \mathbb{P}\left((k\rightsquigarrow W)|(s^{Q}\ni k)\right)\cdot\pi_{k}^{Q},
\end{align*}

\noindent where $(k\rightsquigarrow W)$ denotes the event \textquotedblleft unit $k$ is assigned to week $W$ according to the assignment procedure\textquotedblright.\\

To compute the weekly assignment probability $\mathbb{P}\left((k\rightsquigarrow W)|(s^{Q}\ni k)\right)$ we compose a dataset with the frame and design variables $x^{(q)}_{k}$ used for the assignment for each unit $k$ as well as the weekly assignment $w_{k}$. Since this survey has a rotating sampling design where each unit $k$ retains the assigned week $w_{k}$ when it is firstly selected in the sample, we only include records corresponding to this first selection. In this way, we have a dataset with the empirical results of the allocation procedure. Only units from the past six time reference periods are included in this dataset.\footnote{Six is the number of periods a statistical unit is kept in the rotating sample.}\\

Next using a random forest routine for this classification problem, we get the resulting probabilities computed by the algorithm for each unit $k$ over the whole dataset. These probabilities are indeed the probabilities $\mathbb{P}\left((k\rightsquigarrow W)|(s^{Q}\ni k)\right)$ which we need. Notice that we do not need to split the dataset into train and test, since we are not going to predict any output. We are just \emph{measuring} the probabilities $\mathbb{P}\left((k\rightsquigarrow W)|(s^{Q}\ni k)\right)$ from the empirical data. Overfitting is indeed beneficial because we want to measure the probabilities for this specific dataset.\\

Now, having both the weekly design weights $d_{k}^{W}=1/\pi_{k}^{W}$ and the weekly target variable values $\delta_{k}(C)$ for the weekly respondent sample $r^{W}\subset s^{W}$, we proceed exactly the same way as in the quarterly time scale. We compute the Horvitz-Thompson estimate $\widehat{A}^{W,HT}_{d}(C)=\sum_{k\in s_{d}^{W}}d_{k}^{W}\delta_{k}(C)$ and apply the two-step calibration procedure \citep{SarLun05a} to adjust for non-response and to reduce variance \citep{INE_EPA09a}, so that finally $$\widehat{A}^{W}_{d}(C)=\sum_{k\in s_{d}^{W}}\omega_{ks}^{W}(\mathbf{x})\delta_{k}(C),$$\noindent where $\omega_{ks}^{W}(\mathbf{x})$ stands for the two-step calibrated weekly sampling weight. As marginal totals for calibration we use the monthly version of the original marginal population totals. This is indeed a decision depending on the production environment: should the weekly marginal population totals be available (nowadays uncommon in official statistics), we would use them; instead, the corresponding monthly population figures are used.\\

\begin{figure}[!htb]
    \centering
    \includegraphics[width=0.75\textwidth]{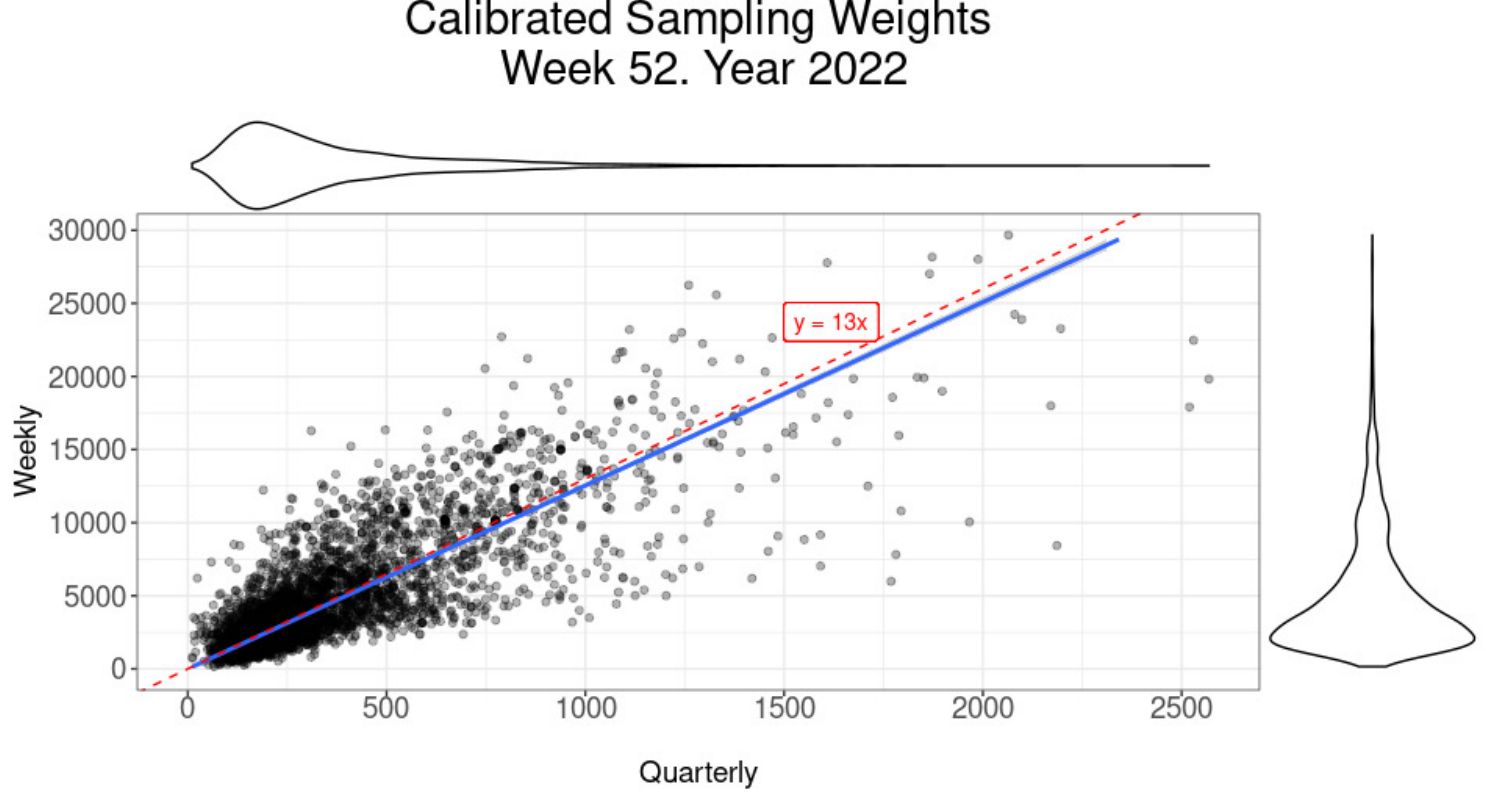}
    \caption{Weekly calibrated sampling weights $\omega_{ks}^{W}(\mathbf{x})$ vs quarterly calibrated sampling weights $\omega_{ks}^{Q}(\mathbf{x})$. The regression line with no intercept  (dashed) has an estimated slope $\hat{\beta}_{1}=12.857$ with $\hat{\sigma}=0.096$.}
    \label{fig:EPA_weekWeights}
\end{figure}

Figure~\ref{fig:EPA_weekWeights} depicts an illustrative example for the 52th week of 2022. We rapidly observe that (i) as expected there roughly exists a factor $13$ to go from the quarterly to the weekly scale and (ii) there exist outliers. Further insight is needed here to understand the model performance (e.g. with the Brier score) and its interrelation with the calibration process and the outlying values.\\

The weekly estimates $\widehat{A}_{d}^{W}(C)$ can now be combined to produce monthly estimates 

\begin{equation}\label{eq:EPA_HT-M}
\widehat{A}_{d}^{M}(C)=\frac{1}{n_{M}}\sum_{W\subset M}\widehat{A}_{d}^{W}(C),
\end{equation}
\noindent where $n_{M}$ denotes the number of weeks $W$ in the month $M$. Notice that this is possible not only for calendrical months $M$.\\

Alternatively, rescuing the aforementioned ergodic hypothesis for month $M$, and computing monthly inclusion probabilities as $\pi_{k}^{M} = \sum_{W\subset M}\pi_{k}^{W}$, we can also produce monthly estimates as $$\widehat{A}_{d}^{M}(C)=\sum_{k\in M}\omega_{ks}^{M}(\mathbf{x})\delta_{k}^{M}(C),$$ where $\omega_{ks}^{M}(\mathbf{x})$ denotes the calibrated sampling weight with initial design weight $d_{k}^{M}=\frac{1}{\pi_{k}^{M}}$ and $\delta_{k}^{M}(C)$ stands for the target variable value assumed for the whole month $M$ from the weekly collected value $\delta_{k}(C)$.\\

\begin{figure}[!htb]
    \centering
    \includegraphics[width=\textwidth]{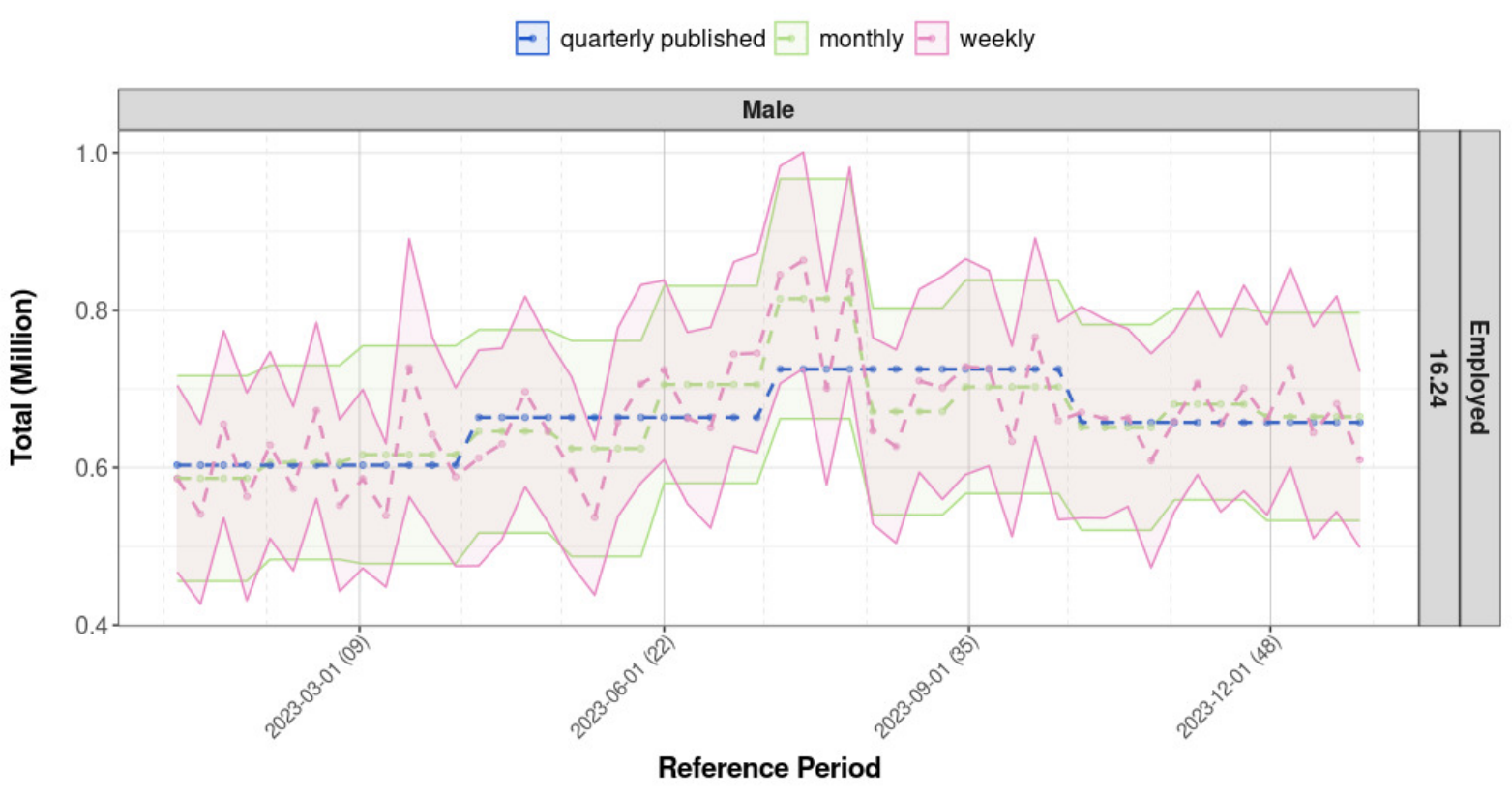}
    \caption{Quarterly, monthly, and weekly estimates for the total number of employed people in the age group 16--24 during 2023. No time-series filtering technique has been applied at this point: sampling variability is clearly visible.}
    \label{fig:EPA_Series}
\end{figure}

In either form, both the calibrated weekly estimates $\widehat{A}^{W}_{d}(C)$ and the monthly estimates $\widehat{A}^{M}_{d}(C)$ show a great variability for the same quarter (in the same flavour as multiple Horvitz-Thompson estimates from the same population, especially for the weekly disaggregation). To reduce this variability we apply a filter to the time series so constructed. This is beyond the methodological scope of this chapter. In Figure \ref{fig:EPA_Series} we show an example of weekly and monthly disaggregations compared to the original quarterly time series before applying any time-series filtering technique. The variability is visible, especially in the weekly scale.\\

This proposal is currently under deeper investigation \citep{EPAMensNTTS2025} to assess different issues potentially impinging on the quality of the final estimates:

\begin{enumerate}
    \item Once higher frequency is gained, accuracy must be our first-order concern. Regarding bias, we remind that probability estimation by random forests is consistent \citep{BiaDevLug08a}, so that $\widehat{A}_{d}^{W,HT}$ is asymptotically unbiased considering both the sampling design and the estimation model. Sampling weights calibration only introduces an asymptotically negligible bias. The size of the training sets (the past six quarters in our case) needs to find a trade-off between the distance in time in the past (so that weekly assignments actually are the result of the current practice, not that of the far past) and the consistency rate.
        \item In the same line, regarding the variance, firstly the estimated variance for the weekly estimator $\widehat{\mathbb{V}}[\widehat{A}_{d}^{W}(C)]$ is computed with the same jackknife methodology as the quarterly time scale \citep{INE_EPA09a}. Then, for the monthly scale we conceive the monthly estimator \eqref{eq:EPA_HT-M} as the result of the sampling design $p(\cdot)$ and a combination of weekly estimates $w$ so that we can write

\begin{align}
    \mathbb{V}[\widehat{A}^{M}_{d}(C)] = \mathbb{V}_{w}[\mathbb{E}_{p}[\widehat{A}^{M}\big|s^{M}]] + \mathbb{E}_{w}[\mathbb{V}[\widehat{A}_{d}^{M}(C)\big|s^{M}]].
\end{align}

This variance can be estimated by

\begin{align}
    \widehat{\mathbb{V}}[\widehat{A}_{d}^{M}(C)] = \widehat{\mathbb{V}}_{JK}[\widehat{A}_{d}^{M}(C)] + \frac{1}{n_{M}}\sum_{W\subset M}\widehat{\mathbb{V}}[\widehat{A}_{d}^{W}(C)],
\end{align}

\noindent where $\widehat{\mathbb{V}}_{JK}[\widehat{A}_{d}^{M}(C)]$ stands for a jacknife estimator of the monthly estimator $\widehat{A}_{d}^{M}(C)$ deleting one week at a time. This rough estimation of the variance needs to be refined takig into account the filtering procedure.

\item We have observed that dropping out the aforementioned ergodic hypothesis on the target variable values during the whole quarter occassionally drives us to three monthly estimates systematically below or above the original quarterly estimates. This needs further investigation and probably the introduction of a benchmarking step to reach full coherence between the original estimates and the time-disaggregated estimates.

\item So far, our focus has been placed on the estimation of population totals. Employment and unemployement rates, thereof derived, must be further estimated and analysed.
\end{enumerate}

\section{Some conclusions}
\label{sec:Concl}

This collection of proposals and pilot experiences allows us to gather some conclusions regarding the use of statistical learning models for the production of official statistics.\\

Our main conclusion is that statistical learning models constitute a versatile tool enabling the improvement of the design, execution, and monitoring of statistical business functions, both traditional functions already in production and novel modifications impinging on different quality dimensions. Thus, in the same spirit as survey methodology has been successfully deployed in production in statistical offices, an adaptation of these organizations to integrate these new methods and technologies must be put in place. This involves different aspects, mainly data, technology, and skills.\\

Machine Learning is a highly data-intensive activity. Therefore, data governance,  data managament, and data architectures are crucial to implement these methods at scale. The amount of pre-processing tasks in our proofs of concept to prepare data and compute regressors (feature engineering) would clearly benefit if a data architecture with fully-fledged metadata is put in place and shared among all surveys. This would reduce the cost for the discovery and development of new ML-based applications.\\

Indeed, in this same line, a repository of regressors or features would be highly advised with a continuously updating process in place incorporating subject matter knowledge. According to our experience, representation learning or feature learning is a crucial step for the performance of the models. The subject matter knowledge needs to be transferred to the models and the identification and construction of features from the available data constitute a fundamental task in this direction.\\

Data quality is another critical aspect for the performance of Machine Learning models. This does not only imply the curation of existing data but also (and even more strategically) the decision about what data to collect and to use for training algorithms. In particular, in relation with the necessary reduction of response burden generated by survey data collection, in our opinion, this high-quality data source should be strategically considered. When planned to be integrated with administrative data and digital transactional data, a change of focus should be undertaken: we need to pursue the quality of the trained models and not the quality of final statistical outputs themselves. The problem of sample selection in statistical offices should progressively focus on a selection problem to assure model quality (either assisted- or dependent-like) not direct estimators quality. In our view, this is already suggesting the role of statistical offices in the new data ecosystems, where statistical knowledge and quality assurance frameworks (e.g.\ for calibration) will be more relevant than statistical products (to be also released in the new future by other actors with more data, higher computational power, and more complex statistical tools -- probably the policymakers themselves \citep[see e.g.][]{GueMar24a}).\\

Data-intensivity requires also computational capacity. This is needed to continuously train the models with new collected and validated data. Technological platforms such as MLOps solutions providing these new functionalities are necessary. This computational environment should be provided with complementary and necessary functionality such as data security, software version control, continuous development and continuous integration tools, model versioning tools, etc.\\

Skills related to statistical learning need to be generalized in the same way as sample survey methodology is general knowledge among production staff in a statistical office. In other words, statistical methodology capabilities should be extended to comprise also statistical learning. For example, while the difference between a Horvitz-Thompson estimator, a ratio estimator, and a GREG estimator is common knowledge among statisticians in a statistical office, this is not the case between a gradient boosting algorithm, a random forest algorithm, and a CART algorithm. Notice that knowledge does not imply that a distributed isolated production organization should be assumed; even when central units specialised on these techniques are considered, the knowledge should be extensive to all production staff.\\

Pilot experiences and proposals need to start their path to implementation in usual production conditions. A standard protocol to promote proofs of concept, minimal viable products, and experimental statistics to official statistics need to be also put in place hopefully following common international guidelines. For example, when should these new features be included (then made compulsory) in national and international legal regulations (e.g. can it be made legally compulsory to deliver a short-term business statistics in, say, 15 days?).\\

Finally, as a prominent suggestion motivated by the high predictive power of these methods in terms of data at statistical units level, we think that the main focus in official statistics should be shifted from the quality in statistical outputs and aggregates as the primary concern of production to the availability of synthetic microdata to increase timeliness, granularity, cost-efficiency, accuracy, and response burden reduction. Once high-quality microdata are in place through a combination of data collection and statistical learning methods, standard aggregation procedures can always be used to produce traditional outputs. Now the challenge is to guarantee the quality of these models and their input training data.\\

\section*{Acknowledgments}

We acknowledge the invaluable collaboration of colleagues from different statistical domains (Structural and Short-Term Business Statistics, Labour Market Statistics, Social Statistics) and the IT department of Statistics Spain (INE). We thank Florian Dumpert for insightful comments on a previous version of this work.

\end{document}